\documentclass[12pt,preprint]{aastex}

\usepackage{natbib}

\newcommand{\GHz}{\mbox{\rm GHz} }

\newcommand{\rms}{ \mbox{\rm rms} }
\newcommand{\mkelvin}{\mbox{\rm mK} }
\newcommand{\ukelvin}{\mbox{$\mu$\rm K}}

\newcommand{\ddeg}         {\mbox{${\rlap.}^\circ$}}
\newcommand{\wmap}{\mbox{\sl WMAP} }

\begin{document}

\title{Seven-Year Wilkinson Microwave Anisotropy Probe (WMAP\altaffilmark{1}) Observations: \\
  Sky Maps, Systematic Errors, and Basic Results }

\author{
{{N. Jarosik}}\altaffilmark{2}, 
{{C. L. Bennett}}\altaffilmark{3}, 
{{J. Dunkley}}\altaffilmark{4}, 
{{B. Gold}}\altaffilmark{3}, 
{{M. R. Greason}}\altaffilmark{5}, 
{{M. Halpern}}\altaffilmark{6}, 
{{R. S. Hill}}\altaffilmark{5}, 
{{G. Hinshaw}}\altaffilmark{7}, 
{{A. Kogut}}\altaffilmark{7}, 
{{E. Komatsu}}\altaffilmark{8}, 
{{D. Larson}}\altaffilmark{3}, 
{{M. Limon}}\altaffilmark{9}, 
{{S. S. Meyer}}\altaffilmark{10}, 
{{M. R. Nolta}}\altaffilmark{11}, 
{{N. Odegard}}\altaffilmark{5}, 
{{L. Page}}\altaffilmark{2}, 
{{K. M. Smith}}\altaffilmark{12}, 
{{D. N. Spergel}}\altaffilmark{12,13},  
{{G. S. Tucker}}\altaffilmark{14}, 
{{J. L. Weiland}}\altaffilmark{5}, 
{{E. Wollack}}\altaffilmark{7}, 
{{E. L. Wright}}\altaffilmark{15}}

\altaffiltext{1}{\wmap\ is the result of a partnership between Princeton
University and NASA's Goddard Space Flight Center. Scientific
guidance is provided by the \wmap\ Science Team.}
\altaffiltext{2}{{Dept. of Physics, Jadwin Hall, %
                    Princeton University, Princeton, NJ 08544-0708}}
\altaffiltext{3}{{Dept. of Physics \& Astronomy, %
                    The Johns Hopkins University, 3400 N. Charles St., %
                    Baltimore, MD  21218-2686}}
\altaffiltext{4}{{Astrophysics, University of Oxford, %
                    Keble Road, Oxford, OX1 3RH, UK}}
\altaffiltext{5}{{ADNET Systems, Inc., %
                    7515 Mission Dr., Suite A100 Lanham, Maryland 20706}}
\altaffiltext{6}{{Dept. of Physics and Astronomy, University of %
                    British Columbia, Vancouver, BC  Canada V6T 1Z1}}
\altaffiltext{7}{{Code 665, NASA/Goddard Space Flight Center, %
                    Greenbelt, MD 20771}}
\altaffiltext{8}{{Univ. of Texas, Austin, Dept. of Astronomy, %
                    2511 Speedway, RLM 15.306, Austin, TX 78712}}
\altaffiltext{9}{{Columbia Astrophysics Laboratory, %
                    550 W. 120th St., Mail Code 5247, New York, NY  10027-6902}}
\altaffiltext{10}{{Depts. of Astrophysics and Physics, KICP and EFI, %
                    University of Chicago, Chicago, IL 60637}}
\altaffiltext{11}{{Canadian Institute for Theoretical Astrophysics, %
                    60 St. George St, University of Toronto, %
                    Toronto, ON  Canada M5S 3H8}}
\altaffiltext{12}{{Dept. of Astrophysical Sciences, %
                    Peyton Hall, Princeton University, Princeton, NJ 08544-1001}}
\altaffiltext{13}{{Princeton Center for Theoretical Physics, %
                    Princeton University, Princeton, NJ 08544}}
\altaffiltext{14}{{Dept. of Physics, Brown University, %
                    182 Hope St., Providence, RI 02912-1843}}
\altaffiltext{15}{{UCLA Physics \& Astronomy, PO Box 951547, %
                    Los Angeles, CA 90095--1547}}


\begin{abstract}
New full sky temperature and polarization maps based on seven years of data from \wmap are presented.
The new results are consistent with previous results, but have improved due to
reduced noise from the additional integration time, improved knowledge
of the instrument performance, and improved data analysis procedures.  The
improvements are described in detail.

 The seven year
data set is well fit by a minimal six-parameter flat $\Lambda$CDM model. The parameters for this model, using the \wmap data
in conjunction with baryon acoustic oscillation data from the Sloan Digital Sky Survey
 and priors on \ensuremath{H_0} from Hubble Space Telescope observations, are:
\ensuremath{\Omega_bh^2} = \ensuremath{0.02260\pm 0.00053}, 
\ensuremath{\Omega_ch^2} = \ensuremath{0.1123\pm 0.0035},
\ensuremath{\Omega_\Lambda} = \ensuremath{0.728^{+ 0.015}_{- 0.016}},
\ensuremath{n_s} = \ensuremath{0.963\pm 0.012}, 
\ensuremath{\tau} = \ensuremath{0.087\pm 0.014} \ and
\ensuremath{\sigma_8} = \ensuremath{0.809\pm 0.024} (68 \% CL uncertainties).  

The temperature power spectrum signal-to-noise ratio per multipole is greater that unity for multipoles 
$\ell \lesssim 919$, allowing a robust measurement of the third acoustic peak. This measurement
results in improved constraints on the matter density, $\ensuremath{\Omega_mh^2} = 
\ensuremath{0.1334^{+ 0.0056}_{- 0.0055}}$, and the epoch of matter-radiation equality,
 $\ensuremath{z_{\rm eq}} = \ensuremath{3196^{+ 134}_{- 133}}$,  
using \wmap ~data alone. 

The new \wmap data, when combined with smaller angular scale microwave background anisotropy data, results in a 3$\sigma$ detection 
of the abundance of primordial Helium, $\ensuremath{Y_{\rm He}} = 0.326 \pm 0.075$.
When combined with additional external data sets, the \wmap data also yield better determinations of the total
mass of neutrinos, $\ensuremath{\sum m_\nu} = \ensuremath{< 0.58\ \mbox{eV}\ \mbox{(95\% CL)}}$, and the 
effective number of neutrino species, $\ensuremath{N_{\rm eff}} = \ensuremath{4.34^{+ 0.86}_{- 0.88}}$.
 The power-law index of the 
primordial power spectrum is now determined to be \ensuremath{n_s} = \ensuremath{0.963\pm 0.012}, 
excluding the Harrison-Zel'dovich-Peebles spectrum by $> 3 \sigma$

These new \wmap measurements provide important tests of Big Bang cosmology.

\end{abstract}

\keywords{cosmic microwave background, space vehicles: instruments}

\section{Introduction}
The Wilkinson Microwave Anisotropy Probe (\wmap) is a NASA sponsored satellite designed to
map the Cosmic Microwave Background (CMB) radiation over the entire sky in five frequency bands.
It was launched in June 2001 from Kennedy Space Flight Center and began surveying the sky 
from its orbit around the  the Earth-Sun L2 point in August 2001. This work and the accompanying
papers comprise the fourth in a series of biennial data releases and incorporates seven years
of observational data. 

Results from the one-year, three-year and five-year results are summarized in \citet{bennett/etal:2003b},
 \citet{jarosik/etal:2007} and
\citet{hinshaw/etal:2009} respectively, and references therein. An overall description of the
 mission including instrument
nomenclature is contained in \citet{bennett/etal:2003} and \citet{limon/etal:prep},  while details of the optical system
 and radiometers can be found in
\citet{page/etal:2003} and \citet{jarosik/etal:2003}. 

The primary data product of \wmap are sets of calibrated sky maps at five frequency bands
centered at 23~\GHz (K band), 33~\GHz (Ka band), 41~\GHz (Q band), 61~\GHz (V band) and 94~\GHz (W band),
including measured noise levels and beam transfer functions that describe the smoothing of the sky signal
resulting from the beam geometries. These maps are provided for Stokes I, Q and U parameters on a year by year 
basis and in a year co-added format, and at several pixel resolutions appropriate for various analyses. Changes
relative to the previous data release include the inclusion of seven years of observational data, a new masking
procedure that simplifies the map-making process, and improvements of the beam maps and window functions.   Details
of the processing used to generate these products are described in the remainder of this work.

In an accompanying paper \citet{gold/etal:prep} utilize the maps in the five frequency bands and
some external data sets to estimate levels of Galactic emission in each map, and describe the generation of 
a set of reduced foreground sky maps based on template cleaning, and a map generated using an Internal 
Linear Combination map (ILC) of \wmap data, both of which are used for analysis of the CMB anisotropy signal.

\citet{larson/etal:prep} describe the measurement of the angular power spectrum of the CMB 
obtained from the reduced foreground sky maps  and the cosmological parameters obtained by fitting 
the CMB power spectra to current cosmological models.

The cosmological implications of the data, including the use of external data sets,  
are discussed by \citet{komatsu/etal:prep}, while \citet{bennett/etal:prep} discuss a number of arguably 
anomalous results detected in previous \wmap data releases.

\citet{weiland/etal:prep} describe the characteristics of a group of point-like objects observed by \wmap 
in the context of their use as microwave calibration sources for astronomical observations.

The remainder of this paper is organized as follows: Section~\ref{sec:dataproc} presents updates on the data processing procedures
used to generate the seven-year sky maps and related data products. Section~\ref{sec:beams} describes  ongoing
efforts to characterize the \wmap beam shapes, while Section~\ref{sec:maps} presents the seven-year sky map data and power spectra, 
describes some additional analyses on the low-$\ell$ polarization power spectra, 
 and summarizes the scientific results obtained from the latest \wmap data set. 

\section{Data Processing Updates} \label{sec:dataproc}
\subsection{Operations}
The sixth and seventh years of \wmap's operation span the interval from 00:00:00 UT, 9 August 2006 
(day number 222) to 00:00:00 UT, 10 August 2008 (day number 222) and includes  nine short periods when 
observations were interrupted. These periods include
eight scheduled events: six station keeping maneuvers, and one maneuver to avoid flying through the 
Earth's shadow (11 November 2007), followed by
a small orbital correction 19 days later. The other interruption to 
observations occurred as a result of the failure of the primary 
transmitter used to telemeter data to Earth.  \wmap was subsequently 
reconfigured to use its backup transmitter and
normal operations resumed with no performance degradation. On 1 August 2008 
(day number 141) \wmap flew through the Moon's 
shadow, causing a $\approx 4\%$ decrease in the incident solar flux lasting $\approx 6$h. \wmap 
remained in normal observing mode throughout this 
event, but data was excluded from sky map processing due to minor instrument thermal perturbations as described 
in \S~\ref{sec:gaps_in_obs}.

\subsection{Calibration}
The algorithm used to calibrate the time ordered data (TOD) is the same as was used 
for the five-year processing \citep{hinshaw/etal:2009}. Calibration occurs 
in two steps. First an hourly absolute 
gain and baseline are determined. The absolute calibration is based on the CMB monopole temperature 
\citep{mather/etal:1999}  and the velocity dependent dipole resulting from \wmap's orbit about the solar system 
barycenter. The calibration is performed iteratively, since removing
the fixed sky signals arising from the barycentric CMB signal and foregrounds requires values of both the gain 
and baseline solutions. The only change made to this step of the calibration procedure is that the initial value
 for the barycentric CMB dipole signal has been updated to agree with the value determined from the five-year analysis.

The  second step in the calibration procedure consists of fitting the hourly radiometer gain values to a  model
which relates the gain to  measured values of radiometer parameters.
The functional form of the gain model is the same as for the five-year data release, but the fitting procedure
now utilizes all seven years of observational data. The seven-year mean changes in the calibration are presented 
in Table~\ref{tab:sigma_0}. Note that the \rms of the fractional calibrations change, $\Delta G/G$, is only 0.05\%. 
The absolute calibration accuracy is estimated to be 0.2\%, unchanged from the previously published value.

\begin{deluxetable}{lcccc}
\tablecaption{ Noise and Calibration Summary for the Template Cleaned and Uncleaned Maps \label{tab:sigma_0}}
\tablehead{ \colhead{DA} & \colhead{$\sigma_0({\rm I})$} & \colhead{ $\sigma_0({\rm Q,U})$}&
\colhead{$\sigma_0(\rm Q,U)$} &\colhead{$\Delta G/G$}\tablenotemark{a}\\ 
& \multicolumn{2}{c}{uncleaned} & template cleaned \tablenotemark{b}& \\
  & \colhead{$(\mkelvin)$} & \colhead{$(\mkelvin)$} & \colhead{$(\mkelvin)$} & \colhead{(\%)}}

\startdata
 K1   &  1.437 & 1.456 & NA & -0.14\\ 
 Ka1  &  1.470 & 1.490 &2.192& -0.01\\
 Q1   &  2.254 & 2.280 &2.741&  0.01\\
 Q2   &  2.140 & 2.164 &2.602  & 0.01\\
 V1   &  3.319 & 3.348 &3.567  &-0.03\\
 V2   &  2.955 & 2.979 &3.174& -0.03\\
 W1   &  5.906 & 5.940 &6.195& -0.05\\
 W2   &  6.572 & 6.612 &6.896& -0.04\\
 W3   &  6.941 & 6.983 &7.283& -0.08\\
 W4   &  6.778 & 6.840 &7.134& -0.12\\
\enddata 
\tablenotetext{a}{$\Delta G / G$ is the change in calibration of the current (seven-year) processing relative to the five-year processing.
A positive value means that features in the seven-year maps are larger than the same features in the five-year map.}
\tablenotetext{b}{The $\sigma_0$ value for the Stokes I template cleaned maps is the same as for the uncleaned maps.}

\end{deluxetable}

\begin{figure}
\epsscale{0.7}
\plotone{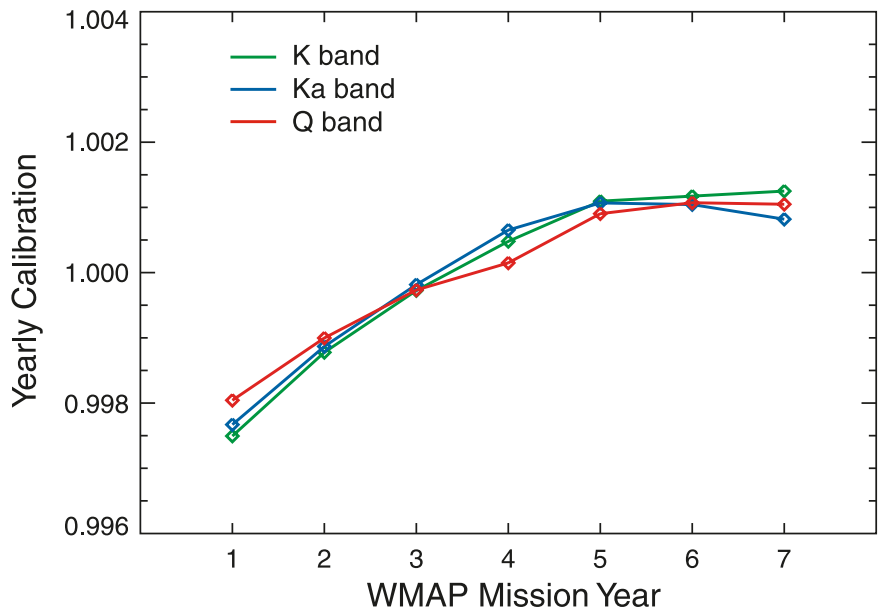}
\caption{Measurements of the year-to-year calibration variation for K, Ka and Q bands obtained by correlating
the Galactic plane signal in the seven-year map to the signal in single year sky maps. Note that the measured 
variations are consistent with the estimated absolute calibration uncertainty of 0.2\%. No significant variation
is seen for the V and W band maps.}
\label{fig:gain_plot}
\end{figure}

For K, Ka, and Q bands, there is a small but significant year-to-year
variation in the \wmap\ calibration that can be seen as variations of
the Galactic plane brightness in yearly sky maps.  This has been
measured by correlating each yearly map against the 7-year map for
pixels at $\vert b \vert < 10\arcdeg$.  A slope and offset is fit to
each correlation and the slope values are adopted as yearly calibration
variations.  These are shown in Figure~\ref{fig:gain_plot} and are consistent with
the 0.2\% absolute calibration uncertainty. 

\subsection{Gaps in Observations} \label{sec:gaps_in_obs}
The \wmap instrument provides the highest quality data when the
observatory is scanning the sky in its nominal observing mode and the instrument thermal 
environment is stable. Periodically, conditions arise that result in one or both of these 
conditions not being met. These situations arise from scheduled station-keeping maneuvers, 
and unscheduled events, such  as solar flares or on-board equipment malfunctions.
Data taken during these periods are excluded from the sky map processing to ensure the highest quality data products.
Previously, potentially corrupted data segments were identified by manually
inspecting the event logs and instrument thermal trend plots. The current data release uses a more objective automated
procedure to identify the unusable data segments. The automated procedure was designed
to approximate the manual procedure, but small differences will occur in the sky coverage between 
the first five years of the current data release and the previous five-year data release. Using the new procedure \wmap
still achieves an overall observing efficiency of $\approx 98.4\%$, down slightly from its previous value of $\approx 99\%$.

The automated procedure identifies suspect events based on the time-derivative of the focal plane assembly (FPA) temperature. 
The measured FPA temperature is averaged over one hour intervals, and the time derivative is formed by 
differencing successive values of these averages. Suspect thermal events are delineated by the times at 
which the magnitude of this derivative initially exceeds then finally falls below a threshold value. 
The threshold has been chosen to be 5 times the \rms deviation of the temperature derivative signal occurring 
during normal observing periods, and corresponds to a value of \mbox{ $\approx 0.75~{\rm mK~h^{-1}}$}.  In situations
 where the observatory was taken out of normal observing mode
(e.g., during a scheduled station keeping maneuver) the duration of the event is lengthened if needed 
to encompass the entire time the observatory was not in normal observing mode. For each event data
from 1.2 hours before the event began to 7.25 hours after the event ended is excluded from sky map 
processing. 

\subsection{Planet Masking}\label{sec:planet_masking}
In the one, three and five-year data analyses, observations when the boresight of each telescope beam fell within $1\ddeg5$ of
Mars, Jupiter, Saturn, Uranus or Neptune were excluded from sky map processing, preventing contamination of 
the sky maps by emission from the planets. Subsequent analysis has shown that even with these exclusion criteria
microwave emission from Jupiter could generate as much as a $75~\ukelvin$ (in K band) errant signal in a narrow 
annulus surrounding the region of excluded observations. Although there is no indication that this influenced the 
cosmological analysis, the radii used to exclude observations have been increased for the seven-year analysis, 
chosen to limit planetary leakage to less than \mbox{1 \ukelvin}. However, given the uncertainty in the determination of 
the beam profiles at these levels, we adopt a conservative upper bound  on possible  planetary signal leakage into
the TOD used for sky map production of $5$ \ukelvin.
 The new values are displayed in Table~\ref{tab:PlanetCuts}.

\begin{deluxetable}{llllll}
\tablecaption{ Data Exclusion Radii for the Planets \label{tab:PlanetCuts}}
\tablehead{
& \multicolumn{5}{c}{Frequency Band}\\
\colhead{Planet}&{K}&{Ka}&{Q}&{V}&{W}}
\startdata
Mars     & $2\ddeg0$ & $1\ddeg5$ & $1\ddeg5$ & $1\ddeg5$ & $1\ddeg5$\\
Jupiter  & $3\ddeg0$ & $2\ddeg5$ & $2\ddeg5$ & $2\ddeg2$ & $2\ddeg0$\\
Saturn   & $2\ddeg0$ & $1\ddeg5$ & $1\ddeg5$ & $1\ddeg5$ & $1\ddeg5$\\
Uranus   & $2\ddeg0$ & $1\ddeg5$ & $1\ddeg5$ & $1\ddeg5$ & $1\ddeg5$\\
Neptune  & $2\ddeg0$ & $1\ddeg5$ & $1\ddeg5$ & $1\ddeg5$ & $1\ddeg5$\\
\enddata 
\end{deluxetable}

\subsection{Expanded Diffuse Galactic Foreground Mask}

The mask for diffuse Galactic emission has been revised by including areas near
the plane where the diffuse foreground cleaning algorithm appears to be less
efficient than for the sky at large. These areas are found by performing a pixel-by-pixel
$\chi^2$ test comparing null maps to cleaned Q-V and V-W maps with resolution degraded from 
r9~\footnote{\wmap uses the HEALPix~\citep{Gorski/etal:2005} pixelization 
and labels resolution as r4, r5, r9 and r10 corresponding to HEALPix $N_{\rm side}$ 
values of 16, 32, 512 and 1024 respectively.} 
 to r5.  All pixels with $\chi^2$ greater
than 4 times that of the $\chi^2$ in the polar caps regions are cut.  Small islands of cut
pixels are eliminated from the cut if they contain 4 or fewer contiguous
pixels.  The two resulting masks based on Q-V and V-W analyses, respectively, are
combined and promoted back to r9.
The edges of the cut are smoothed by convolving the mask with a Gaussian of
$3\arcdeg$ FWHM and cutting the result at a value of 0.5.  As the final step, 
the smoothed cut is combined with the five-year {\it KQ85} or {\it KQ75} cut~\citep{gold/etal:prep}.  
The resulting masks, termed {\it KQ85y7} and {\it KQ75y7},
admit $78.3\%$ and $70.6\%$ of the sky, respectively, as compared to $81.7\%$
and $71.6\%$ for the five-year versions---a decrease of the admitted sky area by
$3.4\%$ of the full sky for {\it KQ85} and $1.0\%$ for {\it KQ75}.  
 
\subsection{Map-Making with Asymmetric Masking} 
The most significant change in the current processing is the use of asymmetric masking in the 
iterative reconstruction of the sky maps. 
\wmap's differential 
design means that the time-ordered-data (TOD) represents differences between the intensities of pairs of points 
on the  sky observed by the two telescope beams. 
Reconstructing sky maps from differential data requires solving a set of linear equations that describe the 
relation between the differential TOD and the sky signal. Solution of this set of equations is performed iteratively, 
using sky maps from earlier iterations to alternately remove 
estimated sky signals for each beam from the TOD, leaving the value associated with the 
opposite beam which is then used to generate the next iteration of the sky map.
This procedure becomes problematic when one of the beams is in a region of high intensity and the 
other is in a region of low intensity. Small errors arising from pixelization, residual 
pointing errors, beam ellipticity or radiometer gain errors limit the degree to which the signal associated with the beam 
in the high emission region can be estimated. Such errors result in errant signals being introduced into 
sky map pixels associated with the beam in the low intensity region. This potential problem is circumvented by
masking portions of the data when such errors are likely to be introduced.

In the \wmap first year results an asymmetric masking procedure was used~\citep{hinshaw/etal:2003d}. 
Asymmetric masking means that when one beam is in a high Galactic 
emission region (as determined by a processing mask~\citep{limon/etal:prep}) and the other beam is in a low Galactic emission region,
only the pixel in the high emission region is iteratively updated. This allows for the reconstruction of full 
sky maps while avoiding the situation that could produce an errant signal. The maps, ${\bf t}_1$, were 
obtained through a simple iterative solution of  the linear equation
\begin{equation}
{\bf t}_{\rm1} = {\bf W} {\bf d}_{\rm 1} \label{eq:pass1_map_eq}
\end{equation} where ${\bf d}_1$ is the pre-whitened time ordered data 
and ${\bf W} = ({\bf M}^T {\bf M})^{-1} \cdot {\bf M} ^{T}$. The matrix ${\bf M}$ is the mapping 
function, which has $N_p$ columns and $N_t$ rows, corresponding to the number of sky map pixels and
time-ordered data points respectively. When the map processing includes polarization degrees of freedom  $N_p$
is four times the number of map pixels, corresponding to the Stokes I, Q and U components and a spurious component, S, used
to absorb effects arising from bandpass differences between the two radiometers comprising each differencing assembly (DA). 
See \citet{jarosik/etal:2007} for a
description of the implementation of the spurious mode, S.    Multiplying a sky map vector by ${\bf M}$ effectively generates 
the time ordered data that would be obtained if \wmap observed a sky corresponding to the map vector.
The sky maps obtained by solving equation~\ref{eq:pass1_map_eq} are an unbiased representation of the true sky signal,
 but do not treat the noise terms optimally.

The \wmap three-year \citep{jarosik/etal:2007} and five-year \citep{hinshaw/etal:2009} analyses generated 
maximum likelihood map solutions using a
conjugate gradient iterative technique. The maximum likelihood estimate of the 
sky map, ${\bf {\tilde t}}$,  is
\begin{equation}
{\bf { \tilde t}} = ({\bf M}^T {\bf N}^{-1} {\bf M})^{-1} \cdot ({\bf M}^T {\bf N}^{-1} {\bf d}) \label{eqn:CGsoln}
\end{equation} 
where ${\bf d}$ is the calibrated but unfiltered time ordered data, and ${\bf N}^{-1}$ is the inverse of 
the radiometer noise covariance matrix. Solution of this equation involves a 
direct evaluation of the last term of equation~\ref{eqn:CGsoln} once, resulting in a map ${\bf t_0} = {\bf M}^T {\bf N}^{-1} {\bf d}$,
followed by the iterative solution of the equation \begin{equation}
{\bf { \tilde t}} = ({\bf M}^T {\bf N}^{-1} {\bf M})^{-1} \cdot {\bf t_0}. \label{eqn:CGsoln2}
\end{equation}
Solving this equation with the conjugate gradient method requires the use of symmetric masking since 
this method can only be used when the matrix multiplying ${\bf t_0}$ (in this case $({\bf M}^T {\bf N}^{-1} {\bf M})^{-1}$) is symmetric . 
Symmetric masking means that if either beam is in a region of high Galactic emission, neither pixel is updated.
Symmetric masking is implemented by simply replacing all occurrences of the 
full sky mapping matrix, ${\bf M}$, with a masked version of the matrix in equations (\ref{eqn:CGsoln} and \ref{eqn:CGsoln2}). Since symmetric 
masking produces sky maps with no information in the high Galactic emission regions, in previous analyses two sets of sky maps were generated,
symmetrically masked maps, and full sky maps incorporating no masking. Data from  full sky maps were used to fill
in regions of the symmetrically masked maps which contained no data. This procedure was a significant advancement over 
the iterative procedure used in the \wmap first year results, in that it produced maximum likelihood maps 
and allowed for 
direct determination of the level of convergence of the solution. Its major disadvantages are that two sets of maps have 
to be generated then pieced together, and there is no simple method of calculating a pixel-pixel 
noise matrix which describes the noise covariance between pixels obtained from the two different input maps. 
The pixel-pixel noise covariance matrix delivered with the data release 
only applied to the pixels in the low-Galactic emission regions, and only diagonal weights are used 
to describe the noise properties in the regions of high Galactic emission.

Incorporating asymmetric masking into the maximum likelihood sky maps solution requires solving the equation 
\begin{equation}
{\bf { \tilde t}} = ({\bf M_{\rm am}^T} {\bf N}^{-1} {\bf M})^{-1} \cdot ({\bf M_{\rm am}^T} {\bf N}^{-1} {\bf d}) \label{eqn:BiCGsoln}
\end{equation}
where ${\bf M}^T_{\rm am}$ is the mapping matrix incorporating  asymmetric masking 
and ${\bf M}$ is the unmasked mapping matrix. Again, this equation is solved by direct 
evaluation of the last term,
\begin{equation}
{\bf t}_0 ={\bf M_{\rm am}^T} {\bf N}^{-1} {\bf d} \label{eqn:BiCG_t0}
\end{equation}
once,  followed by an iterative solution of the equation
\begin{equation}
{\bf { \tilde t}} = ({\bf M}^T_{\rm am} {\bf N}^{-1} {\bf M})^{-1} \cdot {\bf t_0}. \label{eqn:BiCG_map_from_t0}
\end{equation}
Note that the matrix $ {\bf M}^T_{\rm am} {\bf N}^{-1} {\bf M}$ is not symmetric, and therefore this equation 
cannot be solved using 
a simple conjugate gradient algorithm as was done previously. A bi-conjugate 
gradient stabilized method \citep{templates} was chosen to solve this equation since it 
offered good convergence properties and was straightforward to implement, requiring relatively minor changes to 
the existing procedures.  This technique
was verified by reconstructing input maps to numerical precision from simulated noise-free TOD. 
Utilizing asymmetric masking eliminates both of the aforementioned shortcomings of the simple conjugate 
gradient method - each DA only requires a single map solution, and a pixel-pixel inverse noise covariance 
matrix can be generated which describes the noise correlation between all the pixels in the map.

\subsection{Calculation of the pixel-pixel noise covariance matrix, ${\bf \Sigma}$} \label{sec:Sigma_inv_calc}
 The pixel-pixel noise covariance matrix is given by 
	${\bf \Sigma} = \langle{\bf {\tilde t}_{\rm n}} {\bf {\tilde t}}^T_{\rm n}\rangle$, where ${\bf t_n}$ is 
the noise component of a reconstructed sky  map and the 
brackets denote an ensemble average. The time-ordered data, ${\bf d}$, used to generate the map  may be written as the sum of 
a signal term and a noise term
\begin{equation}
{\bf d } = {\bf Mt } + {\bf n}
\end{equation}
where ${\bf M}$ is the full sky (unmasked) mapping matrix, {\bf t} is a vector representing the input sky signal and {\bf n} is 
a vector corresponding
 to the radiometer noise, its covariance being described as
\begin{equation}
	{\bf N} = \langle{\bf n}{\bf n}^{\rm T}\rangle. \label{eqn:TOD_noise}
\end{equation}
The noise component of the maximum likelihood asymmetrically masked map solution may be written as
\begin{equation}
{\bf { \tilde t_{\rm n}}} = ({\bf M_{\rm am}^T} {\bf N}^{-1} {\bf M})^{-1} \cdot ({\bf M_{\rm am}^T} {\bf N}^{-1} {\bf n}). \label{eqn:BiCGnoisesoln}
\end{equation}
The noise covariance matrix becomes
\begin{eqnarray}
{\bf \Sigma}&=&  \langle( {\bf M}^T_{\rm am} {\bf N}^{-1} {\bf M})^{-1} ({\bf M}^T_{\rm am} {\bf N}^{-1} {\bf n}) \cdot 
	      [({\bf M}^T_{\rm am} {\bf N}^{-1} {\bf M})^{-1} ({\bf M}^T_{\rm am} {\bf N}^{-1} {\bf n})]^T \rangle\\
	      &=&  \langle ( {\bf M}^T_{\rm am} {\bf N}^{-1} {\bf M})^{-1} ({\bf M}^T_{\rm am} {\bf N}^{-1} {\bf n}) \cdot 
	      ({\bf n}^T {\bf N}^{-1} {\bf M}_{\rm am})( {\bf M}^T {\bf N}^{-1} {\bf M}_{\rm am})^{-1} \rangle\\
	      &=&  ( {\bf M}^T_{\rm am} {\bf N}^{-1} {\bf M})^{-1} ({\bf M}^T_{\rm am} {\bf N}^{-1})\cdot(\langle{\bf n}
	      {\bf n}^T\rangle {\bf N}^{-1}){\bf M}_{\rm am}( {\bf M}^T {\bf N}^{-1} {\bf M}_{\rm am})^{-1} \label{eqn:noise_cancel}\\
	      &=&  ( {\bf M}^T_{\rm am} {\bf N}^{-1} {\bf M})^{-1} \cdot ({\bf M}^T_{\rm am} {\bf N}^{-1}
	      {\bf M}_{\rm am})\cdot( {\bf M}^T {\bf N}^{-1} {\bf M}_{\rm am})^{-1}.
\end{eqnarray}
In practice the terms ${\bf M}^T_{\rm am} {\bf N}^{-1} {\bf M}$ and ${\bf M}^T_{\rm am} {\bf N}^{-1}
	      {\bf M}_{\rm am}$ are evaluated at r4 with the ${\bf N}^{-1}$ normalized to unity at zero lag, 
and the inverse pixel-pixel noise covariance matrix is generated by forming the  product
\begin{equation}
 {\bf \Sigma}^{-1} =  ( {\bf M}^T {\bf N}^{-1} {\bf M}_{\rm am}) \cdot ({\bf M}^T_{\rm am} {\bf N}^{-1}
	      {\bf M}_{\rm am})^{-1}\cdot( {\bf M}^T_{\rm am} {\bf N}^{-1} {\bf M}). \label{eq:SigmaInv}
\end{equation}
Regions of the ${\bf \Sigma}^{-1}$ matrix corresponding to combinations of Stokes I, Q and U are then converted
to noise units by dividing by the appropriate combinations of $\sigma_0(I)$ and $\sigma_0(Q,U)$ given in Table~\ref{tab:sigma_0}. 

 The term ${\bf M}^T_{\rm am} {\bf N}^{-1}{\bf M}_{\rm am} $ is numerically inverted and is used as the source for 
the off-diagonal terms of the preconditioner for the bi-conjugate gradient stabilized algorithm.

\subsection{$N_{\rm obs}$ Fields of the Maps}\label{sec:Nobs_calc}
The asymmetric masking used to generate sky maps requires a new procedure for calculating the effective number of 
observations, $N_{\rm obs}$, for each map pixel. In previous data releases the effective number of observations was 
calculated by accumulating the number of time ordered data points falling within each pixel,
 weighted by the appropriate transmission imbalance coefficients, $x_{\rm im}$,
and polarization projection factors ($\sin 2\gamma, ~\cos 2\gamma$). These values corresponded to the diagonal elements 
of the inverse pixel-pixel noise covariance matrix ${\bf \Sigma}^{-1} = {\bf M}^T {\bf N}^{-1} {\bf M}$ (\citet{jarosik/etal:2007} 
equation (26)), 
evaluated assuming white radiometer noise, i.e. ${\bf N}^{-1} = {\bf I}$. The simple procedure for evaluating these terms 
is not applicable to the form of ${\bf \Sigma}^{-1}$ for the asymmetrically masked maps, equation~(\ref{eq:SigmaInv}). 
The $N_{\rm obs}$ fields of the r9 and r10 sky maps were generated by evaluating the diagonal elements of 
equation~(\ref{eq:SigmaInv}) with  ${\bf N}^{-1} = {\bf I}$ using sparse matrix techniques.
The $N_{\rm obs}$ fields of the r4 sky maps are described in \S~\ref{sec:r4_map_calc}. 
 
\subsection{Projecting Transmission Imbalance Modes from the ${\bf \Sigma^{-1}}$ Matrices}\label{sec:loss_imb}
As described in \citet{jarosik/etal:2007}, errors in the determination of the transmission imbalance parameters, 
$x_{\rm im}$, are a potential source of systematic artifacts in the reconstructed sky maps. These time independent parameters,
which specify the difference between the transmission between the A-side and B-side optical systems for
each radiometer, are measured from the flight data,  and are presented in Table~\ref{tab:x_im}. 
To prevent biasing the cosmological analyses, sky map modes
that can be excited by these measurement errors are identified and projected from the $\Sigma^{-1}$ matrices.

\begin{deluxetable}{crc|crc}
\tablecaption{ Seven-year Transmission Imbalance Coefficients \label{tab:x_im}}
\tablehead{ \colhead{Radiometer} & \colhead{$x_{\rm im}$} & \colhead{Uncertainty} & \colhead{Radiometer} & 
\colhead{$x_{\rm im}$} & \colhead{Uncertainty}}
\startdata 
       K11 &     -0.00063 &      0.00022 &
       K12 &      0.00539 &      0.00010 \\
      Ka11 &      0.00344 &      0.00017 &
      Ka12 &      0.00153 &      0.00011 \\
       Q11 &      0.00009 &      0.00051 &
       Q12 &      0.00393 &      0.00025 \\
       Q21 &      0.00731 &      0.00058 &
       Q22 &      0.01090 &      0.00116 \\
       V11 &      0.00060 &      0.00025 &
       V12 &      0.00253 &      0.00068 \\
       V21 &      0.00378 &      0.00033 &
       V22 &      0.00331 &      0.00106 \\
       W11 &      0.00924 &      0.00207 &
       W12 &      0.00145 &      0.00046 \\
       W21 &      0.00857 &      0.00227 &
       W22 &      0.01167 &      0.00154 \\ 
       W31 &     -0.00073 &      0.00062 &
       W32 &      0.00465 &      0.00054 \\ 
       W41 &      0.02314 &      0.00461 &
       W42 &      0.02026 &      0.00246 \\

\enddata 
\tablecomments{Transmission imbalance coefficients, $x_{\rm im}$, and their uncertainties determined from 
the seven-year observational data.}
\end{deluxetable}

The procedure follows the same  method as used in the three and five years analyses and consists of generating a one-year
span of simulated time ordered data using the nominal values of the loss imbalance parameters. Sky maps are processed
from this archive using the input values of $x_{\rm im}$ and altered values to simulate errors in the measured values 
of the coefficients. The differences between the resultant maps are used to identify map modes
resulting from processing the data with altered $x_{\rm im}$ values. Two modifications have been made to this procedure relative
to the previous analyses: (1) The simulated sky maps are formed at r4 by direct evaluation of equations (\ref{eqn:BiCG_t0},
\ref{eqn:BiCG_map_from_t0}) at r4. This change was adopted to eliminate contamination of the transmission imbalance templates
by the poorly measured sky map modes associated with monopoles in the Stokes I and 
spurious mode S sky maps. It was found that varying 
the $x_{\rm im}$ factors produced large changes in the amplitudes of these poorly measured modes in addition 
to exciting the mode associated with the loss imbalance. Through utilization of a singular value decomposition 
it was possible to null the poorly measured modes associated with the
aforementioned monopoles while preserving the modes associated with the transmission imbalance while 
evaluating equation (\ref{eqn:BiCG_map_from_t0}).
(2) Transmission imbalance modes were evaluated both for the case when 
the $x_{\rm im}$ for both radiometers comprising each DA were increased 20\%
above their measured values, and for the case when the $x_{\rm im}$ value for radiometer 1 was increased by 10\% and 
that of radiometer 2 decreased
by 10\%. (Previously the only combination used was that in which both $x_{\rm im}$ values were increased.)  
For six of the DAs very similar sky map modes were generated by both 
sets of $x_{\rm im}$, while for the V2, W1, W2, and W4
somewhat different modes were generated for the two different sets. As a result, 
only one mode was projected from the K1, Ka1, Q1, Q2, and W3 
matrices while two modes were projected out of the the V2, W1, W2 and W4  matrices. In each case the modes were removed 
from ${\bf \Sigma^{-1}} $following the method described in \citet{jarosik/etal:2007}.

\subsection {Low Resolution Single Year Map Generation} \label{sec:r4_map_calc} 
The low resolution sky maps were generated by performing an inverse 
noise weighted degradation of the high resolution maps that takes into account the 
intra-pixel noise correlations of the high resolution input maps . The weight matrix for the polarization (Stokes Q and U) 
and spurious mode (S) of each high resolution map pixel, p9, is given by 
\begin{equation}
{\bf N}_{\rm obs}({\rm p9}) = \left( \begin{array}{ccc}
    N_{\rm QQ} & N_{\rm QU} & N_{\rm QS}\\
    N_{\rm QU} & N_{\rm UU} & N_{\rm SU}\\
    N_{\rm QS} & N_{\rm SU} & N_{\rm SS}
\end{array} \right ),
\end{equation}
where each element $ N_{\rm XY}$ is an element of ${\bf \Sigma}^{-1}$ (equation~\ref{eq:SigmaInv}) evaluated 
as described in \S~\ref{sec:Nobs_calc}. For each DA year combination
the Q, U, S map sets were generated as 
\begin{eqnarray}
\left (   \begin{array}{c}
  Q_{\rm p4} \\
  U_{\rm p4} \\
  S_{\rm p4}
\end{array} \right )& = & \left( {\bf N}_{\rm obs}^{\rm tot}({\rm p4}) \right)^{-1}
\sum_{{\rm p9} \in {\rm p4}}{\bf N}_{\rm obs}({\rm p9}) \left (   \begin{array}{c}
  Q_{\rm p9} \\
  U_{\rm p9} \\
  S_{\rm p9}
\end{array} \right ),\\
{\bf N}_{\rm obs}^{\rm tot}({\rm p4})& = &\sum_{{\rm p9} \in p4} {\bf N}_{\rm obs}({\rm p9}).
\end{eqnarray}
The $N_{\rm obs}$ fields of the low resolution sky maps contain the diagonal elements of the 
corresponding portions of the ${\bf \Sigma}^{-1}$ matrices as described in \S~\ref{sec:Sigma_inv_calc} before the 
scaling by $\sigma_0$ is applied. 

\subsection {Low Resolution Multi-year Map Generation} \label{sec:r4_map_calc_7yr} 
The single year  single DA maps were combined to form a seven-year map for each DA and seven-year maps for 
each frequency band. The individual low resolution maps were inverse noise weighted using the
${\bf \tilde{\Sigma}}$ matrices (\S~\ref{sec:loss_imb}), scaled to reflect the
noise level of each DA. The weighted polarization maps were formed as

\begin{eqnarray}
\left (   \begin{array}{c}
  Q \\
  U 
\end{array} \right )& = &  {\bf \tilde{\Sigma}}^{\rm tot}
\sum_{\rm yr, DA} {\bf \tilde{\Sigma}^{-1}}_{\rm yr, DA} \left (   \begin{array}{c}
  Q \\
  U \\
\end{array} \right )_{\rm yr, DA},\\
{\bf \tilde{\Sigma}}^{\rm tot}& = &\left(\sum_{\rm yr, DA}   {\bf \tilde{\Sigma}^{-1}}_{\rm yr, DA} \right)^{-},
\end{eqnarray}
where $ ()^{-}$ represents a pseudo-inverse of the sum in parenthesis.  This summation is performed over 
the year/DA combinations to be included in the final map.
For K and Ka-bands the pseudo-inverse is calculated by performing a singular value decomposition of the sum of the 
$QU \times QU$ ${\bf \tilde{\Sigma}^{-1}}$ matrices and inverting all its eigenvalues except for the smallest,
which is set to zero. This procedure de-weights the one nearly singular mode associated with the monopoles in 
the I and S maps. No modes are removed from the Q, V and W-band maps. 

\section{Beam Maps and Window Functions} \label{sec:beams}

The seven-year \wmap beams and window functions result from a
refinement of the data reduction methods used for the five-year beams,
with no major changes in the processing steps.

Briefly, in the five-year analysis,
the beam maps were accumulated from
TOD samples with Jupiter in either the A or the B side beam.  Sky
background was subtracted using the seven-year full-sky maps, with the
band dependent Jupiter exclusion radii (\S\ref{sec:planet_masking}).  Because of
the motion of Jupiter, and the numerous Jupiter observing seasons\footnote{A ``season'' is a $\approx 45$ day period when a planet falls within the scan pattern of
\wmap. Each planet typically has two seasons per year. }, the sky coverage of the maps was $100\%$ in
spite of this masking.

As far as possible, the
beam transfer functions $b_\ell$ in Fourier space were computed
directly from Jupiter data. However, at low signal levels, some information was incorporated
from beam models.
The A and B side Jupiter data were fitted
separately by a physical optics model comprising feed horns with
fixed profiles, together with primary and
secondary mirrors described by fit parameters.
The geometrical configuration of these components was fixed.  
The fitted parameters described small distortions of the 
mirror surface shapes.

The beam models were combined with the Jupiter data at low signal
levels by a hybridization algorithm.  This process was optimized
to yield the minimum uncertainty in beam solid angle, given the
instrumental noise, under the conservative assumption that the systematic
error in the beam model was 100\%.  Details, as well as a summary of
beam processing in earlier data releases, were given by
\citet{hill/etal:2009}.

In addition to the use of three more
seasons of Jupiter data, the following refinements, 
described below in more detail,  have been made for the seven-year beam processing: 

\begin{itemize}
\item Subtraction of background flux from Jupiter observations is
improved by 
expanding the exclusion radius for sky maps (\S~\ref{sec:planet_masking})
\item The beam modeling procedure was modified by adding additional terms to the 
description of the
secondary mirror figure to approximate a hypothetical tip/tilt.
\item The rate of convergence of the physical optics models
is improved by correcting an orthogonalization error in the
conjugate gradient code.
\item The physical optics fit is driven harder in two senses:
\begin{enumerate}
\item A smaller change in $\chi^2$ between conjugate gradient descent
steps is required to declare convergence.
\item The model is fitted over a wider field of view in the observed
beam map.
\end{enumerate}
\end{itemize}

The change in beam profiles and transfer functions between the five-
and seven-year analyses is conveniently summarized by a comparison
of solid angles for the 10 DAs, given in Table
\ref{tab:beams:omega}.  The solid angle increases of 0.8\% in K1
and 0.4\% in Ka1 are the result of the improved background estimates
for the Jupiter observations.  The solid angle changes
for the Q1--W4 DAs have multiple causes.  First, the instrumental
noise in the Jupiter samples is different in detail.  Second, the
beam models have been refitted and differ slightly from the five-year
versions.  Third, the increased S/N  of the Jupiter data in
the beam wings means that for some
DAs, the hybrid threshold is optimized to a lower gain with 7
years of data. This value is the limit below which observed
intensity values in the Jupiter TOD are replaced with computed
values from the beam model.  For the 5 year data analysis, the hybrid thresholds
were 3, 4, 6, 8, and 11 dBi, respectively, for bands K, Ka, Q, V,
and W \citep{hill/etal:2009}.  For the seven-year data analysis, the V and W thresholds are
7 and 10 dBi, respectively, while those for K, Ka, and Q are
unchanged.

As shown in the
table, the aggregate solid angle changes for V and W, which are the bands 
used in the high-$\ell$ TT power spectrum, are $\sim 0.1\%$ or less.
Table \ref{tab:beams:omega} also gives the current values of the 
forward gain $G_m$ and of the factor $\Gamma_{\mathrm{ff}}$ for converting
antenna temperature to flux in Jy.  New hybrid beam profiles and beam
transfer functions $b_\ell$ are available online from 
the Legacy Archive for Microwave Background Data Analysis (LAMBDA).
\begin{deluxetable}{cccccccc}
  \tabletypesize{\scriptsize}
  \tablewidth{0pt}
  \tablecolumns{5}
  \tablecaption{\wmap\ Seven-year Main Beam Parameters
    \label{tab:beams:omega}}
  \tablehead{
    \colhead{} &
    \colhead{$\Omega_{\mathrm{7yr}}^S$\tablenotemark{a}} & 
    \colhead{$\Delta(\Omega_{\mathrm{7yr}}^S)/\Omega^S$\tablenotemark{b}} & 
    \colhead{$\frac{\Omega_{\mathrm{7yr}}^S}{\Omega_{\mathrm{5yr}}^S}-1$\tablenotemark{c}} &
    \colhead{$G_m$\tablenotemark{d}} &
    \colhead{$\nu_{\mathrm{eff}}^{\mathrm{ff}}$} &
    \colhead{$\Omega_{\mathrm{eff}}^{\mathrm{ff}}$} &
    \colhead{$\Gamma_{\mathrm{\mathrm{ff}}}$\tablenotemark{e}} \\
    \colhead{DA} & 
    \colhead{(sr)} &
    \colhead{(\%)} &
    \colhead{(\%)} &
    \colhead{(dBi)} &
    \colhead{(\GHz)}&
    \colhead{(sr)}&
    \colhead{($\mu$K Jy$^{-1}$)}}
  \startdata
  \cutinhead{For 10 Maps}
  K1  & $2.466\times10^{-4}$ & 0.6 &  0.8 & 47.07 &22.72 &$2.519\times10^{-4}$ & 250.3 \\
  Ka1 & $1.442\times10^{-4}$ & 0.5 &  0.4 & 49.40 &32.98 &$1.464\times10^{-4}$ & 204.5 \\
  Q1  & $8.832\times10^{-5}$ & 0.6 & -0.1 & 51.53 &40.77 &$8.952\times10^{-5}$ & 218.8 \\
  Q2  & $9.123\times10^{-5}$ & 0.5 & -0.2 & 51.39 &40.56 &$9.244\times10^{-5}$ & 214.1 \\
  V1  & $4.170\times10^{-5}$ & 0.4 &  0.0 & 54.79 &60.12 &$4.232\times10^{-5}$ & 212.8 \\
  V2  & $4.234\times10^{-5}$ & 0.4 & -0.1 & 54.72 &61.00 &$4.281\times10^{-5}$ & 204.4 \\
  W1  & $2.042\times10^{-5}$ & 0.4 &  0.2 & 57.89 &92.87 &$2.044\times10^{-5}$ & 184.6 \\
  W2  & $2.200\times10^{-5}$ & 0.5 & -0.3 & 57.57 &93.43 &$2.200\times10^{-5}$ & 169.5 \\
  W3  & $2.139\times10^{-5}$ & 0.5 & -0.5 & 57.69 &92.44 &$2.139\times10^{-5}$ & 179.0 \\
  W4  & $2.007\times10^{-5}$ & 0.5 &  0.5 & 57.97 &93.22 &$2.010\times10^{-5}$ & 186.4 \\
  \cutinhead{For 5 Maps}
  K   & $2.466\times10^{-4}$ & 0.6 &  0.8 & 47.07 &22.72 &$2.519\times10^{-4}$ & 250.3 \\
  Ka  & $1.442\times10^{-4}$ & 0.5 &  0.4 & 49.40 &32.98 &$1.464\times10^{-4}$ & 204.5 \\
  Q   & $8.978\times10^{-5}$ & 0.6 & -0.2 & 51.46 &40.66 &$9.098\times10^{-5}$ & 216.4 \\
  V   & $4.202\times10^{-5}$ & 0.4 & -0.1 & 54.76 &60.56 &$4.256\times10^{-5}$ & 208.6 \\
  W   & $2.097\times10^{-5}$ & 0.5 &  0.0 & 57.78 &92.99 &$2.098\times10^{-5}$ & 179.3 \\
  \enddata
  \tablenotetext{a}{Solid angle in azimuthally symmetrized beam.}
  \tablenotetext{b}{Relative error in $\Omega^S$.}
  \tablenotetext{c}{Relative change in $\Omega^S$ between five-year and seven-year analyses.}
  \tablenotetext{d}{Forward gain $=$ maximum of gain relative to isotropic,
    defined as $4\pi/\Omega^S$.  Values of $G_m$ in Table 2 of Hill et al.
    (2009) were taken from the physical optics model, rather than
    computed from the solid angle, and therefore do not obey this
    relation.}
  \tablenotetext{e}{Conversion factor to obtain flux density from
    the peak \wmap\ antenna temperature, for a free-free spectrum with $\beta=-2.1$.
    Uncertainies in these factors are estimated as 0.6, 0.5, 0.6, 0.5 and 0.7\% for 
    K, Ka, Q, V and W band DAs respectively.}
 
\end{deluxetable}
\subsection{Seven-year Beam Model Fitting \label{sec:beammodel}}

\subsubsection{Secondary Tip/Tilt}

One change in the seven-year beam computations is to introduce approximate
tip/tilt terms for the secondary mirror into the physical optics
model. This change is motivated by the stable $\sim 0\fdg1$ offset in the
collective B-side boresight pointings, as compared to pre-flight
expectations \citep{hill/etal:2009}.  In previous beam fits, this
offset was absorbed by two free-floating nuisance parameters.  

The introduction of secondary tip/tilt is a heuristic attempt
to improve the fidelity of the beam model.  Because a displacement in
beam pointings could result from various small mechanical
displacements in the instrument, or from a combination of them, the
beams contain too little information to support a mechanical analysis.
We emphasize that the boresight pointings on both the A and the B
sides are equally stable since launch, to $<10\arcsec$
\citep{jarosik/etal:2007}, an estimate that is confirmed by the 
full seven-year Jupiter data.

For convenience, we approximate tip/tilt as a purely planar surface
distortion of the secondary mirror, in addition to the Bessel modes
used previously.  The pivot line is allowed to vary
by including a term for scalar displacement together with tip/tilt
slopes in two orthogonal directions.

This parameter subspace has been explored using Monte Carlo beam
simulations, which reveal a strong degeneracy between the scalar
offset and a tilt in the spacecraft $YZ$ plane, because both
displacements move the illumination spot on the primary mirror in the
up or down direction.  Therefore, we have investigated a family of
solutions by adopting an \emph{Ansatz} for the mechanical constraints,
such that (1) one edge of the secondary is allowed to pivot while
pinned at the designed position, (2) the opposite edge undergoes a
small displacement, and (3) the secondary mirror as a whole is rigid.

These constraints result in two discrete solutions for each secondary
mirror; for each side of the instrument, the solution requiring the
smaller mechanical displacement is chosen.  The resulting edge
displacement is $-4.4$ mm on the B side, which corresponds to a
hypothetical motion of one edge of the secondary mirror away from the
secondary backing structure and toward the primary.  Similarly, this
method results in a hypothetical displacement of one edge of the
A-side secondary by $0.2$ mm away from the primary.  However, as
already explained, this mechanical interpretation is uncertain.
For subsequent stages of beam modeling, the distortion coefficients
mimicking tip/tilt and displacement of the secondary mirrors are held fixed.

\subsubsection{Fitting Process}

The overall fitting of beams progresses much as it did for the five-year
data.  On each of the A and B sides, the primary mirror is modeled
with Fourier modes added to the nominal mirror figure.  Similarly, the
secondary mirror is modeled with Bessel modes.  A complete fit is done
using modes with spatial frequency on the mirror surface up to some
maximum value.  When convergence is achieved, the current best-fit
parameters become the starting point for a new fit with a higher
spatial frequency limit.  The primary mirror spatial frequencies are
indexed using an integer $k$, such that the wavelength of the mirror
surface distortion is $280\ \textrm{cm}/k$.  The secondary modes
are specified by the Bessel mode indexes $n$ and $k$.
Details are given in \citet{hill/etal:2009}, \S 2.2.2.
The maximum spatial frequencies fitted are defined by $k_\mathrm{max}=24$ on
the primary and $n_\mathrm{max} = 2$ on the secondary.

Because the Fourier modes are non-orthogonal over the circular domain 
of the primary mirror distortions, the fitting is
actually done in an orthogonalized space \citep{hill/etal:2009}.  A bug in the
orthogonalization code was corrected before the seven-year analysis began; 
this bug only affected speed of convergence, not the definition of
$\chi^2$, so the five-year results remain valid.  However, this
change made it practical to refit the beam models \emph{ab
initio} from the nominal mirror figures, i.e., with zero for all
Fourier and Bessel coefficients, rather than beginning from the five-year
solution.  

The $\chi^2$ changes from the five-year to the seven-year beam models are
shown in Table \ref{tab:beams:chisq}.  The $\chi^2$ values shown for
the five-year models are recomputed using seven-year data.
It is clear from Table \ref{tab:beams:chisq}
that the overall improvement in the model for the 10 DAs collectively
is driven primarily by the W band on the A side and the V and W bands
on the B side.

\begin{figure}
\plotone{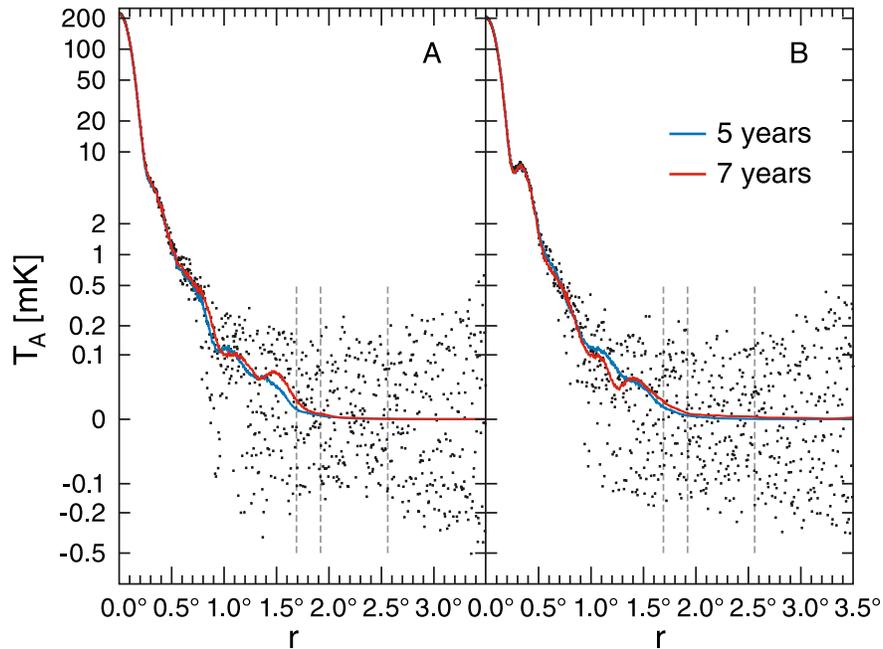}
\caption{Physical optics beam models for the W1 DA on the A side
(left) and B side (right) of the \wmap instrument.
The ordinate is scaled by a $\sinh^{-1}$ function that provides a
smooth transition between linear and logarithmic regimes.
\emph{Blue}: five-year models; \emph{red}: seven-year models.
\emph{Points}: seven-year
Jupiter beam data averaged in radial bins of $\Delta r = 0\farcm5$.
\emph{Dashed lines}: Radii at which hybrid beam profiles consist
of 90\%, 50\%, and 10\% Jupiter data, respectively, from smaller
to larger radii.  Model differences inside $r\sim1\fdg7$ are mostly
suppressed in the hybrid beam profiles, whereas model differences outside
$r\sim2\fdg6$ are mostly retained.
\label{fig:beams:profile}}
\end{figure}

Figure \ref{fig:beams:profile} shows model beam profiles from an
example DA, W1, from both the five-year and the seven-year analyses.  The
A-side model has a similar profile in the five-year and seven-year fits, with
small trade-offs in fit quality at various radii.  However, the B-side
model shows a new feature in the seven-year fitting, namely, an elevated
response at large radii and very low signal levels in the V and W
bands.  This feature is seen at radii larger than $2\arcdeg$ in the
right-hand panel of Figure \ref{fig:beams:profile}, where the seven-year
profile (red) is very slightly higher than the five-year profile (blue).  The five-year
and seven-year profiles for the example DA differ by a factor of $\sim
2-10$ in this region, although both are $\gtrsim 50$ dB down from the
beam peak.

Several attempts have been made using jumps in the fitting parameters
(as in simulated annealing) to find a region of parameter space in the
A-side fit that would lead to the existence of elevated tails similar
to those on the B side.  So far, such attempts have failed.  The best
A-side fit lacks the elevated-tail feature, while the best B-side fit
has it.

Just as in the five-year analysis, none of the beam models meets a formal
criterion for a good fit, since the number of degrees of freedom is of
order $10^5$, or $10^4$ per DA, while the overall $\chi_\nu^2$ is
$\sim 1.05$.  However, the model beams are used only in a restricted
role, more than $\sim 45$ dB below the beam peak, so that the primary
contribution to the beams and $b_\ell$ is directly from Jupiter data.  As
a result, the effect of any particular feature of the models is either
omitted or diluted in the final result.  However, the B-side model tail
mentioned above increases the beam uncertainty slightly at large
radii as compared to the five-year analysis.

\begin{figure}
\plotone{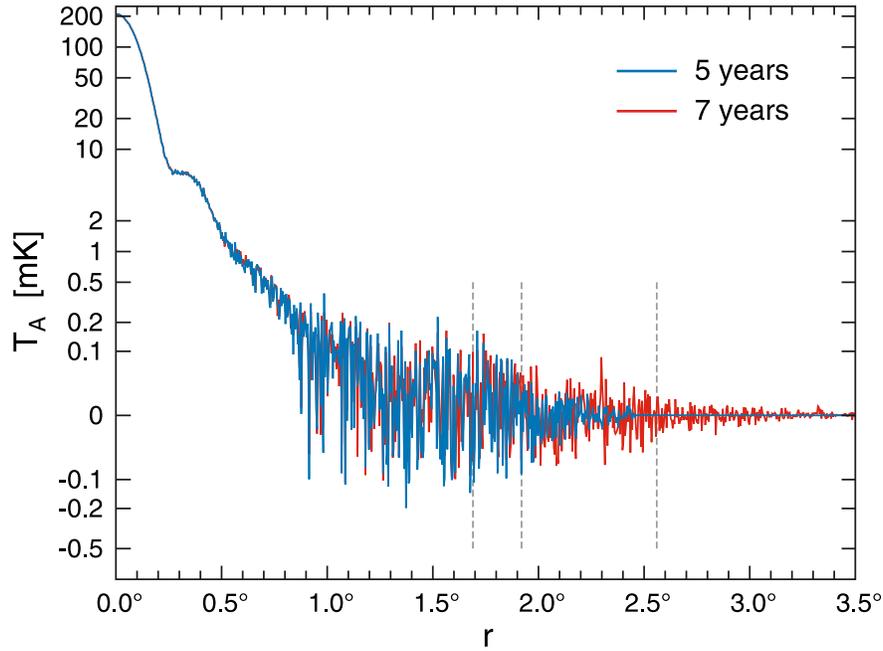}
\caption{W1 hybrid beam profiles from five-year (blue) and seven-year (red) analysis,
combined for the A and B sides.
The ordinate is scaled by a $\sinh^{-1}$ function that provides a
smooth transition between linear and logarithmic regimes.
\emph{Dashed lines}: Radii at which hybrid beam profiles consist
of 90\%, 50\%, and 10\% Jupiter data, respectively, from smaller
to larger radii.  The noise shows that use of Jupiter data extends 
effectively to larger
radii in the seven-year analysis.
\label{fig:beams:hybrid}}
\end{figure}

The process for incorporating information from the models into the
Jupiter-based beams is explained by \citet{hill/etal:2009}.  This
process is termed \emph{hybridization}.  Briefly,
the beam profiles are integrated from Jupiter TOD.  The
two-dimensional coordinates of each TOD point within the beam model
are determined.  If the model gain is below a certain threshold, then
the model gain is substituted for the Jupiter measurement.  The
thresholds are optimized for each DA by minimizing the uncertainty in
the solid angle, under the assumption that the beam model is subject
to a scaling uncertainty of $100\%$.  Final hybrid profiles combining
the A and B sides are shown for the W1 DA, for both the five-year and
the seven-year analysis, in Figure \ref{fig:beams:hybrid}.

\begin{figure}
\begin{center}
\includegraphics[height=6.5in]{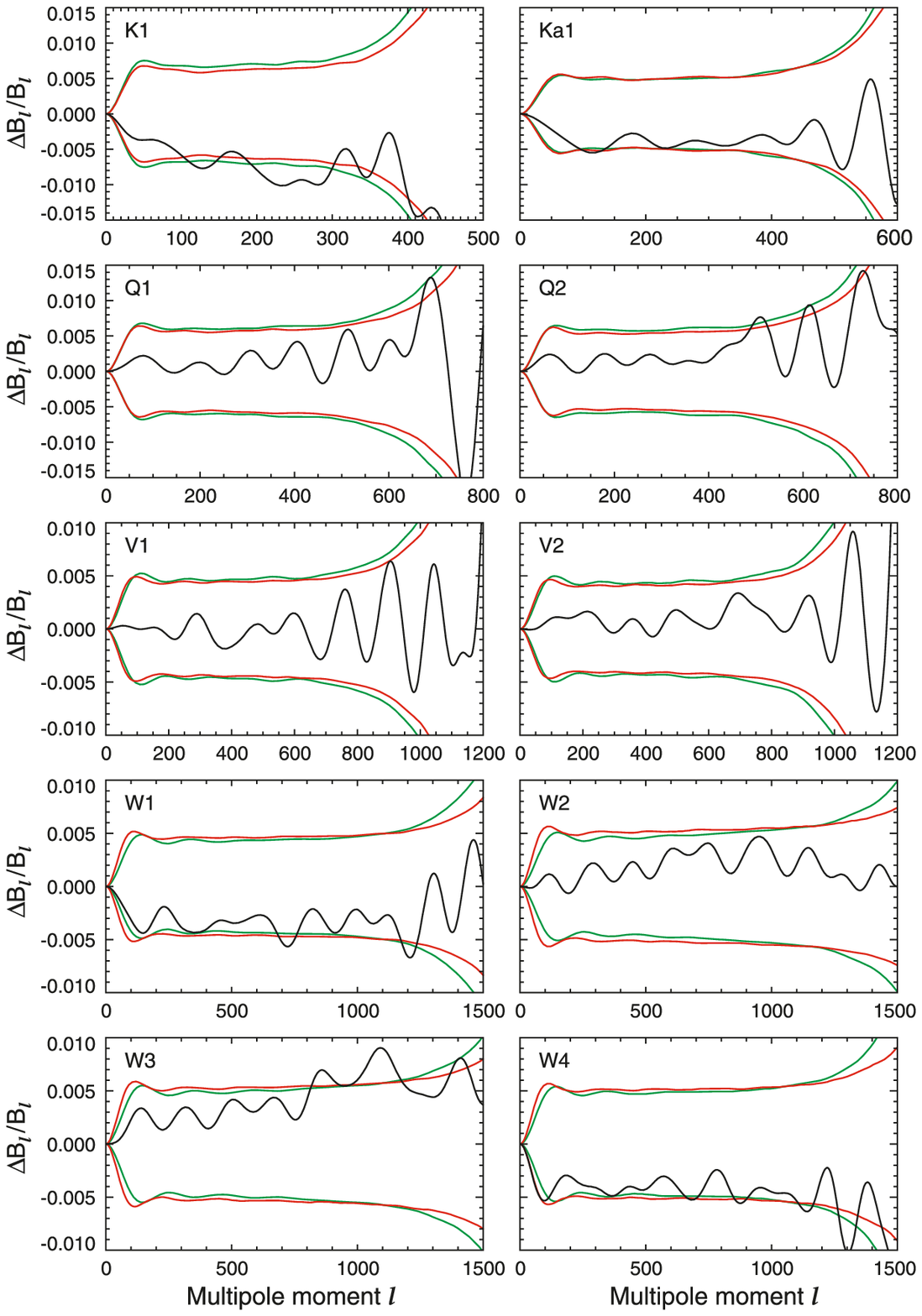}
\end{center}
\caption{ \footnotesize Comparison of beam transfer functions
and uncertainties between five-year and seven-year analyses.
\emph{Black}: Relative change
in beam transfer functions from five to seven years in the sense 
$(\textrm{7 yr}-\textrm{5 yr})/(\textrm{5 yr})$.
\emph{Green}:  five-year $1\sigma$ error envelope.  \emph{Red}:
seven-year $1\sigma$ error envelope.  The seven-year $b_\ell$ are
largely within $1\sigma$ of the five-year $b_\ell$, while the change
in the error envelope itself is small.  In W band, modeling differences 
between the A and B sides introduce an increase in the uncertainty plateau
for multipoles $\ell \lesssim 1000$, whereas the small angle (high
$\ell$) uncertainty is decreased for all bands.
\label{fig:beams:bl}}
\end{figure}

The beam transfer function, $b_\ell$, is integrated directly from the
hybrid beam profiles, and the error envelope for $b_\ell$ is computed
using a Monte Carlo method.  These procedures are the same as in the
five-year analysis \citep{hill/etal:2009}.  A comparison of five-year and
seven-year $b_\ell$ and the corresponding uncertainties is shown in 
Figure \ref{fig:beams:bl}.

\begin{deluxetable}{cccc}
  \tabletypesize{\footnotesize}
  \tablewidth{0pt}
  \tablecolumns{4}
  \tablecaption{Changes in $\chi^2$ from five-year to seven-year beam fits
    \label{tab:beams:chisq}}
  \tablehead{
    \colhead{} &
    \colhead{$\chi_\nu^2$} &
    \colhead{$\chi_\nu^2$} &
    \colhead{$\Delta\chi_\nu^2$} \\
    \colhead{DA} & 
    \colhead{(5 year)} &
    \colhead{(7 year)} &
    \colhead{}}
  \startdata
  \cutinhead{Side A}
  all & 1.060 & 1.049 & -0.011 \\
  K1  & 1.009 & 1.006 & -0.004 \\
  Ka1 & 1.012 & 1.016 &  0.004 \\
  Q1  & 1.078 & 1.074 & -0.003 \\
  Q2  & 1.094 & 1.102 &  0.008 \\
  V1  & 1.144 & 1.155 &  0.011 \\
  V2  & 1.162 & 1.171 &  0.008 \\
  W1  & 1.225 & 1.171 & -0.053 \\
  W2  & 1.239 & 1.151 & -0.088 \\
  W3  & 1.244 & 1.181 & -0.063 \\
  W4  & 1.208 & 1.156 & -0.052 \\
  \cutinhead{Side B}
  all & 1.065 & 1.047 & -0.018 \\
  K1  & 1.022 & 1.017 & -0.005 \\
  Ka1 & 1.010 & 1.009 & -0.000 \\
  Q1  & 1.093 & 1.046 & -0.046 \\
  Q2  & 1.045 & 1.052 &  0.007 \\
  V1  & 1.288 & 1.171 & -0.117 \\
  V2  & 1.178 & 1.159 & -0.019 \\
  W1  & 1.226 & 1.197 & -0.029 \\
  W2  & 1.169 & 1.155 & -0.013 \\
  W3  & 1.223 & 1.203 & -0.021 \\
  W4  & 1.226 & 1.165 & -0.061 \\
  \enddata
\end{deluxetable}

\subsection{Flux Conversion Factors and Beam Solid Angles for Point Sources}
The conversion factor from peak antenna temperature to flux density for a point source is given by
\citep{page/etal:2003b}

\begin{equation}
\Gamma = c2/2 k_B \Omega_{eff} (\nu_{eff})^2
\end{equation}
where $\Omega_{eff}$ is the effective beam solid angle of the A and B sides combined
and $\nu_{eff}$ is the effective band center frequency, both of which
depend on the source spectrum.  New values of these quantities
have been calculated that supersede previous results given for point sources.

For a point source with antenna temperature spectrum $T_A \propto \nu^{\beta}$, the
effective frequency is determined from
\begin{equation}
\nu_{eff}^{\beta} = \int f(\nu) G_m(\nu) \nu^{\beta} d\nu / \int f(\nu) G_m(\nu) d\nu
\end{equation}
where $f(\nu)$ is the passband response, and $G_m(\nu)$ is the forward gain.
This is consistent with the definition used by \citet{jarosik/etal:2003} for a
beam-filling source, except in that case the forward gain is not included.
Values of $\nu_{eff}(\beta)$ are calculated using pre-flight bandpass measurements
and pre-flight GEMAC\footnote{Goddard Electromagnetic Anechoic Chamber} measurements of forward gain.  A correction for
scattering is applied to the GEMAC measurements,
\begin{equation} 
G_m(\nu) = G_m^{GEMAC}(\nu) e^{(-(4 \pi \sigma / \lambda)^2)}
\end{equation}
where $\sigma$ is an effective rms primary mirror deformation whose value is set
separately for each DA based on the degree of scattering estimated from the pass 4
physical optics modeling.

The effective solid angle is calculated as
\begin{equation}
\Omega_{eff}  = \Omega(Jupiter)  \Omega_{GEMAC}(\beta)/\Omega_{GEMAC}(Jupiter) 
\end{equation}
where $\Omega(Jupiter)$ is the pass 4 beam solid angle determined from Jupiter observations
and $\Omega(GEMAC)$ is the beam solid angle calculated for a given source spectrum using
the scattering-corrected GEMAC measurements.  From eq (25) of \citet{page/etal:2003},

\begin{equation}
\Omega(GEMAC) = 4 \pi \int f(\nu) \nu^{\beta} d\nu / \int f(\nu) G_m(\nu) \nu^{\beta} d\nu
\end{equation}

Values of $\nu_{eff}$, $\Omega_{eff}$, and $\Gamma$ for a point source with $\beta=-2.1$,
typical for the sources in the WMAP point source catalog, are given in Table 2.

\section{Sky Map Data and Analysis}\label{sec:maps}
\begin{figure}
\epsscale{.7}
\plotone{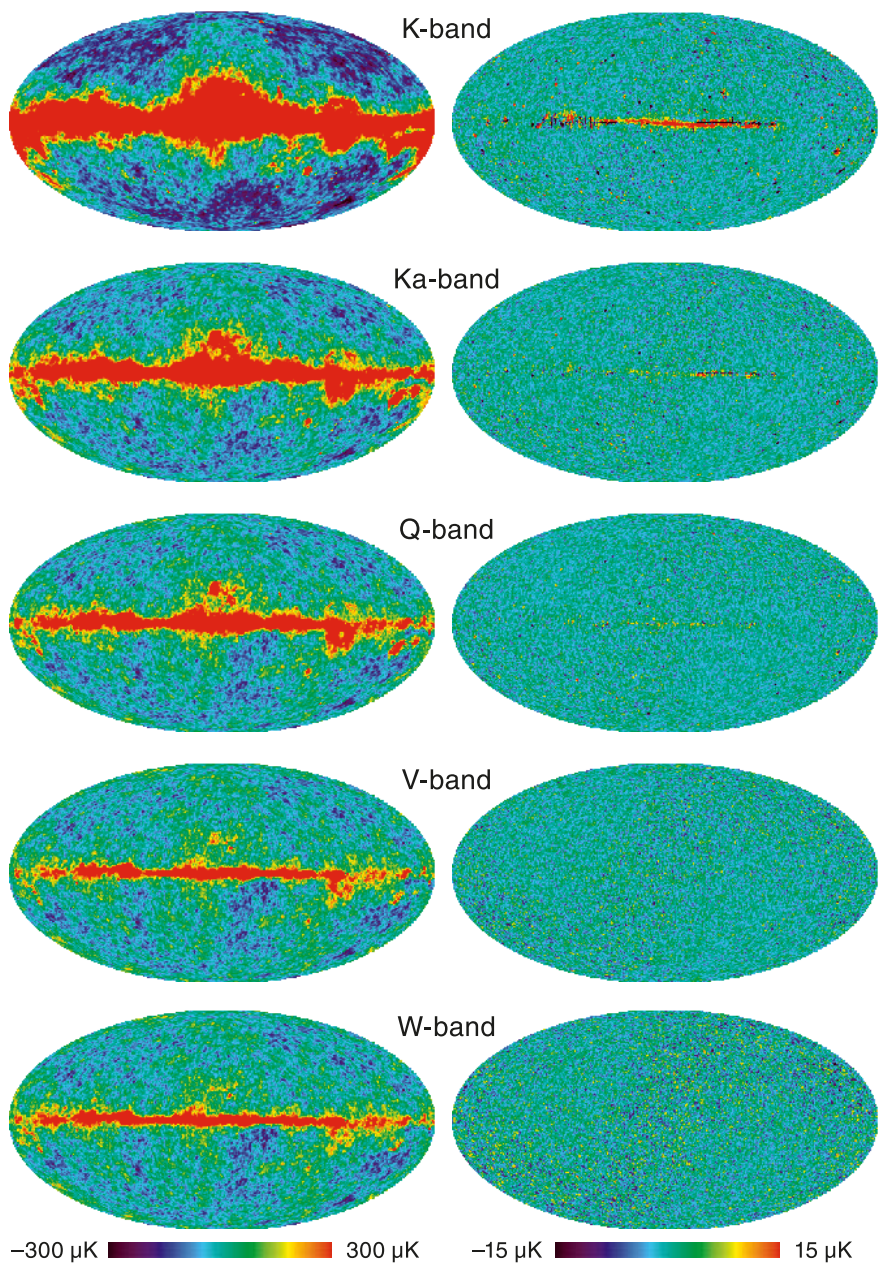}
\caption{\footnotesize Plots of the Stokes I maps in Galactic coordinates. The left column displays the seven-year average
maps, all of which have a common dipole signal removed. The right column displays the difference 
between the seven-year average maps and the previously published five-year average maps, 
adjusted to take into account the slightly
different dipoles subtracted in the seven-year and five-year analyses and the slightly differing calibrations.
 All maps have been smoothed with 
a $1^{\circ}$ FWHM Gaussian kernel. The small Galactic plane signal in the difference maps arises from the
difference in calibration ($0.1\%$) and beam symmetrization between the five-year and seven-year processing. Note that the
temperature scale has been expanded by a factor of 20 for the difference maps.  \label{fig:i_diff_maps}}
\end{figure}

The seven-year sky maps are consistent with the 5 year maps apart from small effects related to the new
processing methods. Figure~\ref{fig:i_diff_maps} displays the seven-year band average Stokes I maps and
the differences between these maps and the published five-year maps. The difference maps have been adjusted
to compensate for the slightly different gain calibrations and dipole signals used in the different analysis.
The small Galactic plane features in the K, Ka and Q band difference maps arise from the slightly different 
calibrations and small changes in the effective beam shapes. Pixels in the Galactic plane region are observed
over a slightly smaller range of azimuthal beam orientation in the current data processing relative to 
previous analyses, resulting in slightly less azimuthal averaging and hence a slightly altered effective beam 
shape. The calibration gain changes relative to the five-year data release are small, all 
below 0.15\% as indicated in Table~\ref{tab:sigma_0}.

\subsection{ The Temperature and Polarization Power Spectra}
\subsubsection{ The Temperature Dipole and Quadrupole }
 
The five-year CMB dipole value was obtained using a Gibbs sampling method~\citep{hinshaw/etal:2009} to estimate 
the dipole signal in both the five-year Internal Linear Combination map (ILC)
and  foreground reduced maps. The uncertainties on the measured parameters were set to encompass both results as
an estimate of the effect of residual foreground signals. The dipole measured from the seven-year data shows 
no significant changes from that obtained from the five-year data, so the best fit dipole parameters remain
unchanged from the five-year values, and are presented in Table~\ref{tab:dipole}. 

\begin{deluxetable}{llllll}
\tablecaption{\wmap Seven-Year CMB Dipole Parameters \label{tab:dipole}}
\tablehead{\colhead{$d_x$\tablenotemark{a}} & \colhead{$d_y$} & \colhead{$d_z$} & \colhead{$d$\tablenotemark{b}} & \colhead{\it l} & \colhead{\it b} \\
\colhead{(mK)}&\colhead{(mK)}&\colhead{(mK)}&\colhead{(mK)}&\colhead{($^\circ)$}&\colhead{($^\circ$)}}

\startdata
$-0.233 \pm 0.005$ & $-2.222\pm 0.004$ & $2.504 \pm 0.003 $&$ 3.355 \pm 0.008 $&$ 263.99 \pm 0.14 $&$ 48.26 \pm 0.03$
\enddata
\tablecomments{The measured values of the CMB dipole signal. These values are unchanged from the five-year values
and are reproduced here for completeness.}
\tablenotetext{a}{ Cartesian components are give in Galactic coordinates. The listed uncertainties include the
effects of noise, masking and residual foreground contamination. The 0.2\% absolute calibration uncertainty
should be added to these values in quadrature.} 
\tablenotetext{b}{The spherical components of the CMB dipole are given in Galactic coordinates and already include
all uncertainty estimates, including the 0.2\% absolute calibration uncertainty.}
\end{deluxetable}

The maximum likelihood value of magnitude of the CMB quadrupole is $l(l+1)C_l/2\pi = 197^{+2972}_{-155}\ukelvin^2~(95\%~CL)$ 
\citep{larson/etal:prep} based on analysis of the ILC map using the {\it KQ85} mask. This value is essentially unchanged from 
the five-year results and lies below the most likely value predicted by the best fit $\Lambda CDM$ model. This value, however,
is not particularly unlikely given the distribution of values predicted by the model, as described in \citet{bennett/etal:prep}.

\begin{figure}
\epsscale{.8}
\plotone{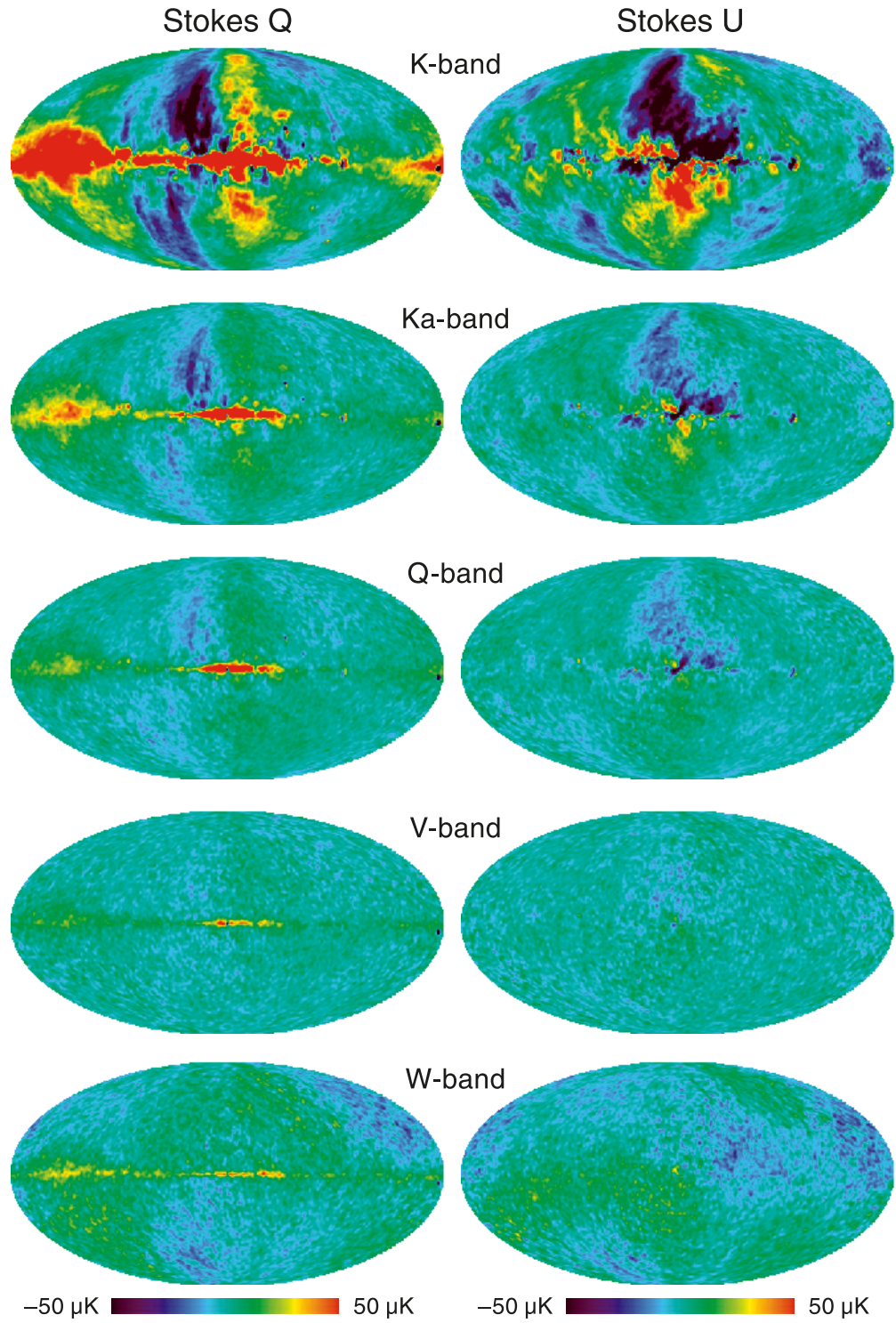}
\caption{\footnotesize Plots of the Seven-year average Stokes Q and U maps in Galactic coordinates.
All maps have been smoothed with a $2^{\circ}$ FWHM Gaussian kernel.  \label{fig:pol_maps}}
\end{figure}

Figure~\ref{fig:pol_maps} displays the seven-year band average maps for Stokes Q and U components for all 5 
\wmap frequency bands.
Polarized Galactic emission is evident in all frequency bands. The smooth large angular scale features 
visible in the W-band
maps, and to a lesser degree in the V-band maps, are the result of a pair of modes that are poorly constrained by the 
map-making procedure. While these dominate the appearance of the map, they are properly de-weighted when these maps are analyzed
using their corresponding ${\bf \Sigma^{-1}}$ matrices, so useful polarization power spectra may be obtained from these maps.
The relatively large amplitudes of these modes limits the utility of using difference plots between the five-year and seven-year map
sets to test for consistency. 

\subsubsection{ Low-$\ell$ W-band Polarization Spectra}
Previous analyses~\citep{hinshaw/etal:2009, Page/etal:2007} exhibited unexplained artifacts in the low-$\ell$ W-band 
polarization power spectra demonstrating an incomplete understanding of the signal and/or noise proprieties of 
these spectra. Specifically the value of $C_{\ell}^{\mathrm{EE}}$ for $\ell = 7$ measured in W-band was found to be significantly higher 
than could  be accommodated by the best fit power spectra, given the  measurement uncertainty.
This result, and several other anomalies, caused these data to be excluded from cosmological analyses.
Significant effort has been expended trying to 
understand these spectra with the goal of eventually allowing their use in cosmological analyses.

A set of null spectra were formed based on the latest uncleaned W-band polarization sky maps to 
test for year-to-year and
DA-to-DA consistency. Polarization cross power spectra were calculated for pairs of maps using the Master
algorithm \citep{hivon/etal:2002} utilizing the full ${\bf \Sigma^{-1}}$ covariance matrix to weight the input maps.
The polarization analysis mask was applied by marginalizing the ${\bf \Sigma^{-1}}$ over pixels excluded 
by the mask to minimize foreground contamination. Appropriately weighted null signal combinations of these spectra were formed
to determine if any individual years or DAs possessed peculiar characteristics. The uncertainties on the power spectra
were evaluated using the Fisher matrix technique \citep{Page/etal:2007} and measured map noise levels.
Since the input sky maps contain both signal 
(mostly of Galactic origin) and noise, an additional term was added to the  
Fisher matrix noise estimate to account for the signal $\times$
noise cross term. The signal component of this term was estimated using the seven-year average power spectra 
values for the combined W1, W2, and W3 DAs. This term was only added for 
multipole/polarization combinations (EE or BB) for which the
estimated signal was greater than  0. 

Table~\ref{tab:cl_chi_sq} shows the result of this analysis. 
The reduced $\chi^2$ combinations in the top panel are evaluated for
data combinations of  all 4 W-band DAs that compare individual years 
to the average of the remaining years. Polarization
combinations EE and BB were evaluated for multipoles ranging from 
2-32. Additional combinations were formed to compare the
first three years of data to the latter four, and to compare data taken 
in odd and even numbered years. All combinations resulted
in reasonable $\chi^2$ and probability to exceed values.

\begin{deluxetable}{ccccc}  
\tabletypesize{\footnotesize}
\tablecaption{ Seven-year Spectrum $\chi^2$ W-band Null Tests  \label{tab:cl_chi_sq}}
\tablehead{ \colhead{Data Combinations } & \colhead{$\chi^2_{\rm EE}$} & \colhead{$PTE_{\rm EE}$} & \colhead{$\chi^2_{\rm BB}$} & \colhead{$PTE_{\rm BB}$}}
\startdata 
\cutinhead{ For DAs W1, W2, W3 and W4}
      \{yr1\} - \{y2, y3, y4, y5, y6, y7\} & 0.944 & 0.56 & 0.976 & 0.50\\
      \{yr2\} - \{y1, y3, y4, y5, y6, y7\} & 0.800 & 0.78 & 0.956 & 0.54\\
      \{yr3\} - \{y1, y2, y4, y5, y6, y7\} & 0.934 & 0.57 & 1.166 & 0.24\\
      \{yr4\} - \{y1, y2, y3, y5, y6, y7\} & 0.850 & 0.70 & 0.920 & 0.59\\
      \{yr5\} - \{y1, y2, y3, y4, y6, y7\} & 0.737 & 0.85 & 1.286 & 0.13\\
      \{yr6\} - \{y1, y2, y3, y4, y5, y7\} & 0.769 & 0.82 & 1.101 & 0.32\\
      \{yr7\} - \{y1, y2, y3, y4, y5, y6\} & 1.108 & 0.31 & 0.831 & 0.73\\
      \{yr1, yr2, y3\} - \{y4, y5, y6, y7\} & 0.608 & 0.96 & 0.98 & 0.49 \\
      \{yr1, yr3, y5, y7\} -  \{y2, y4, y6\} & 0.860 & 0.69 & 1.06 & 0.38 \\
\cutinhead{For 7 years of data}
      \{W1\} -  \{W2, W3, W4\} & 1.195 & 0.21 & 0.982 & 0.49 \\
       \{W2\} -  \{W1, W3, W4\} & 0.802 & 0.77 & 0.926 & 0.58 \\
       \{W3\} -  \{W1, W2, W4\} & 0.890 & 0.64 & 0.897 & 0.63 \\
       \{W4\} -  \{W1, W2, W3\} & 1.163 & 0.24 & 2.061 & 0.0005 \\
       \{W1, W2\} -  \{W3, W4\} & 0.867 & 0.68 & 1.361 & 0.09 \\
       \{W1, W3\} -  \{W2, W4\} & 0.866 & 0.68 & 1.219 & 0.19 \\
       
\enddata 
\tablecomments{ $\chi^2$ tests for various null combinations of low-$\ell$ Master polarization power spectra obtained 
from uncleaned W-band sky maps. Reduced $\chi^2$ values are presented along with the probability to exceed (PTE) values based on 31 
degrees of freedom, corresponding to $2 \leq \ell \leq 32$. Polarization cross power spectra were obtained for each DA year $\times$ 
DA year combination, then
appropriately weighted combinations generated to null any sky signal. Predicted uncertainties were obtained using the standard Fisher matrix
formalism incorporating the inverse noise covariance matrices and the measured sky map noise levels. Since individual power spectra estimates
do include signals (mostly foreground), the uncertainties include a contribution for the signal $\times$ noise cross term as explained
in the text. The only anomalous point occurs when the seven-year W4 data is compared to the seven-year data from the remaining W-band DAs.}
\end{deluxetable}

The lower half of the table contains the results of a similar analysis, but in this case combinations were formed to isolate individual DAs.
The W4 is singled out with a reduced $\chi^2$ value of 1.163 which has only a probability to exceed of only 0.05\%. This DA has an 
unusually large 
1/f knee frequency, which makes it particularly susceptible to systematic artifacts. 
For this reason it is excluded from the analysis below, but
continues to be studied in case  it might contain clues as to the nature of the 
low-$\ell$ polarization anomalies seen in the other W-band DAs.  

\begin{figure}
\epsscale{0.65}
\plotone{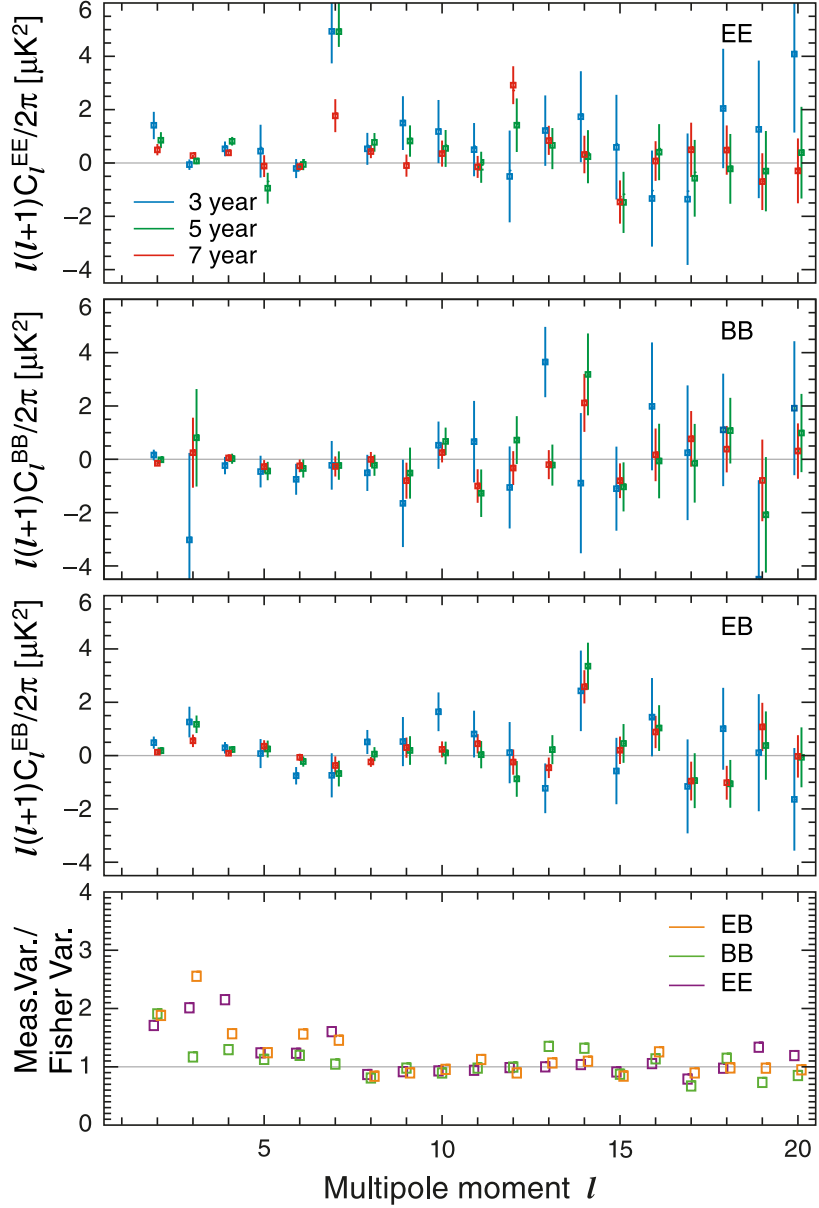}
\caption{Low $\ell$ Master polarization cross power spectra from the first three, five, and seven years 
of data from the W1, W2 and W3  foregrounds reduced polarization sky maps~\citep{gold/etal:prep} for the first three, 
five and seven years of observations. These three time ranges
contain 36, 105 and 210 individual cross spectra respectively. The $\ell$ values 
for the different time ranges have been offset for clarity. The top three panels contain the Master power spectra and error bars
based on a Fisher matrix analysis. The bottom panel is the ratio of the measured variance between the individual power
spectra estimates ( DA  year $\times$ DA year ) and the variance predicted by the Fisher matrix calculation using the
full $\Sigma^{-1}$ inverse noise covariance matrix. Note the good agreement
for $\ell  \geq 8$. \label{fig:kdpol_spectra}}
\end{figure}

Figure~\ref{fig:kdpol_spectra} displays polarization power spectra for the EE, BB and EB modes for the first three, five, and seven years of 
template cleaned individual year maps from the W1, W2 and W3 DAs.
These spectra were also obtained using the Master algorithm  utilizing  the mask-marginalized ${\bf \Sigma^{-1}}$ covariance matrix 
sky map weighting. Only cross power spectra are included, so the measurement noise does not 
bias the measured values. The uncertainties are based on the Fisher matrix
technique, and have not had the signal $\times$ noise term added, since any signal remaining in the cleaned maps is expected to be small. 
(Master algorithm based low-$\ell$ power spectra are not used in any cosmological analyses.)

\begin{figure}
\epsscale{0.65}
\plotone{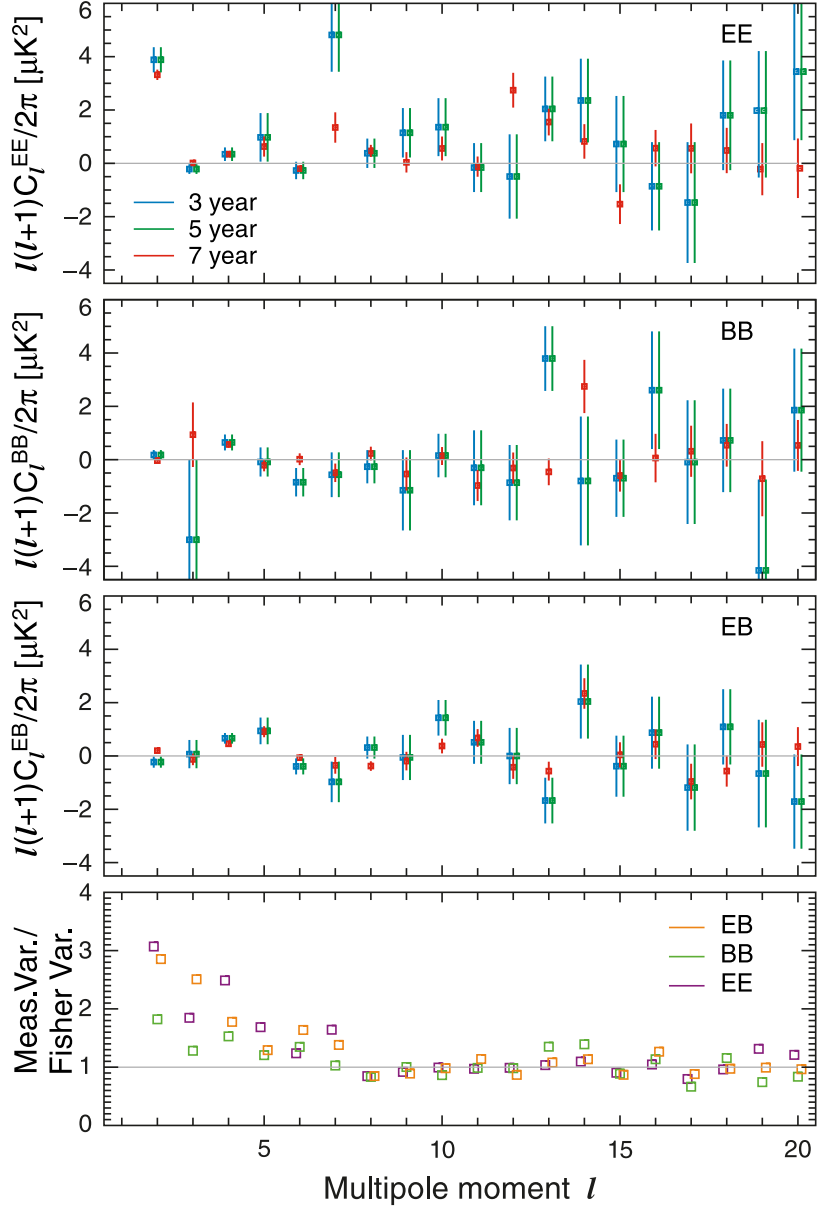}
\caption{Low $\ell$ Master polarization cross power spectra from the first three, five, and seven years 
of data from the W1, W2 and W3 uncleaned sky maps. The $\ell$ values have been offset slightly for the different
time ranges for clarity. The bottom panel is the ratio of the measured variance between the individual power
spectra estimates ( DA  year $\times$ DA year ) and the variance predicted from a Fisher matrix calculation using the
full $\Sigma^{-1}$ inverse noise covariance matrix. Note the good agreement
for $\ell  \geq 8$. \label{fig:uncleaned_spectra}}
\end{figure}

There is generally good agreement between the values for the three different time ranges. Note that the high value seen in the 
three and five-year analysis for $C_{7}^{\mathrm{EE}}$ has fallen significantly with the additional two years of data. In fact, had the data
been take in reverse order (starting with year seven) the $C_{7}^{\mathrm{EE}}$ value would not have been identified 
as particularly anomalous. However, the 
$C_{12}^{\mathrm{EE}}$ value has risen in the seven-year combination. To see if these signals might be artifacts of
the foreground cleaning, a similar analysis
was performed on the uncleaned sky maps and is displayed in Figure~\ref{fig:uncleaned_spectra}. By comparing 
the two figures it can be seen that the
foreground cleaning mainly affects the multipoles $\ell \leq 5$ leaving the $\ell = 7, 12$ multipoles relatively unchanged.
 
The bottom panel of these figures plot the ratio between the variance measured
among the individual cross spectra for each multipole/polarization combination, 
to that predicted by the Fisher matrix technique. For $\ell \geq 8$
there is good agreement between the measured variance and analytical estimate, but the measured variance slowly grows larger with decreasing
$\ell$ for $\ell \leq 7$ for both the cleaned and uncleaned sky maps. The Fisher matrix noise estimate utilizes the 
${\bf \Sigma^{-1}}$ matrix that describes the noise correlation between the
pixels of the Stokes Q and U maps. This matrix incorporates information regarding the sky 
scanning pattern, through the mapping matrices ${\bf M}$ and
${\bf M_{\rm am}}$, and correlations in the radiometer noise, through the  use of inverse radiometer noise covariance matrix ${\bf N^{-1}}$. This
method does not, however, include the effect of any noise correlations that might be 
introduced through the baseline fitting and calibration
procedure. A program was undertaken to determine if the baseline fitting and calibration 
procedure could be affecting the low-$\ell$ polarization 
results and is described in the following section.

\subsubsection{Calibration Induced Spectral Uncertainties}
Uncertainties in the polarization power spectra arising from calibration 
of the TOD  were evaluated using the Fisher information matrix method . 
New sets of noise matrices were calculated including degrees of freedom associated
with the gain and baseline parameters  used in the calibration procedure. The values of these
two parameters were approximated as piecewise functions, each assumed constant over one hour
intervals.
The calibration procedure was modeled as a $\chi^2$ minimization of the residuals
between the TOD and a predicted TOD calculated by applying the gain and
baseline values for each time interval to a model signal consisting of CMB anisotropy and
a time dependent dipole due to \wmap's motion about the solar system barycenter. Fisher 
information matrices were calculated for this
model and inverted to form noise matrices. Uncertainties in the recovered low-$\ell$
power spectra were calculated using this method by marginalizing over the values of the
gain and baseline parameters. A similar calculation was performed with the gain and baseline
terms omitted. The effects of the calibration on the uncertainties in the recovered power
spectra were measured by comparing the results of these two calculations.

As expected, the difference in the predicted uncertainties was small for all but the very
lowest multipoles. This occurs because the  signal from the low-$\ell$ multipoles enters the TOD
on the longest time scales, and therefore is most affected by a the calibration procedure which
fits the low frequency dipole signal used for calibration. For E-mode polarization
the maximum increase in uncertainty in $a_{lm}$ is about 3.5\% occurring for $\ell = 3$ and is less
than 1\% for other multipoles. For the B-mode $a_{lm}$  uncertainties for $\ell = 2, 3$ increase
by 40\% and 90\% respectively,  while the effect on other multipoles is less that 1\%.
The $\ell = 3$ B-mode polarization was already known to be very poorly measured by \wmap, since its symmetry,
combined with \wmap's geometry and scan pattern, generate extremely long period signals
(periods exceeding 10 minutes) in the TOD. The fact that the calculation described above
correctly identified this mode supports the validity of the methodology used, but the overall
results do no fully explain the excess variance observed in all the W-band $\ell \leq 7$ polarization
multipoles. The low-$\ell$ W-band polarization data therefore continue to be excluded from cosmological analysis. 
However, there is no evidence suggesting any
compromise of the high-$\ell$ W-band polarization data, so it it now included 
in evaluation of the high-$\ell$ TE power spectrum.

\subsection{ Science Highlights }
\begin{figure}
\epsscale{0.9}
\plotone{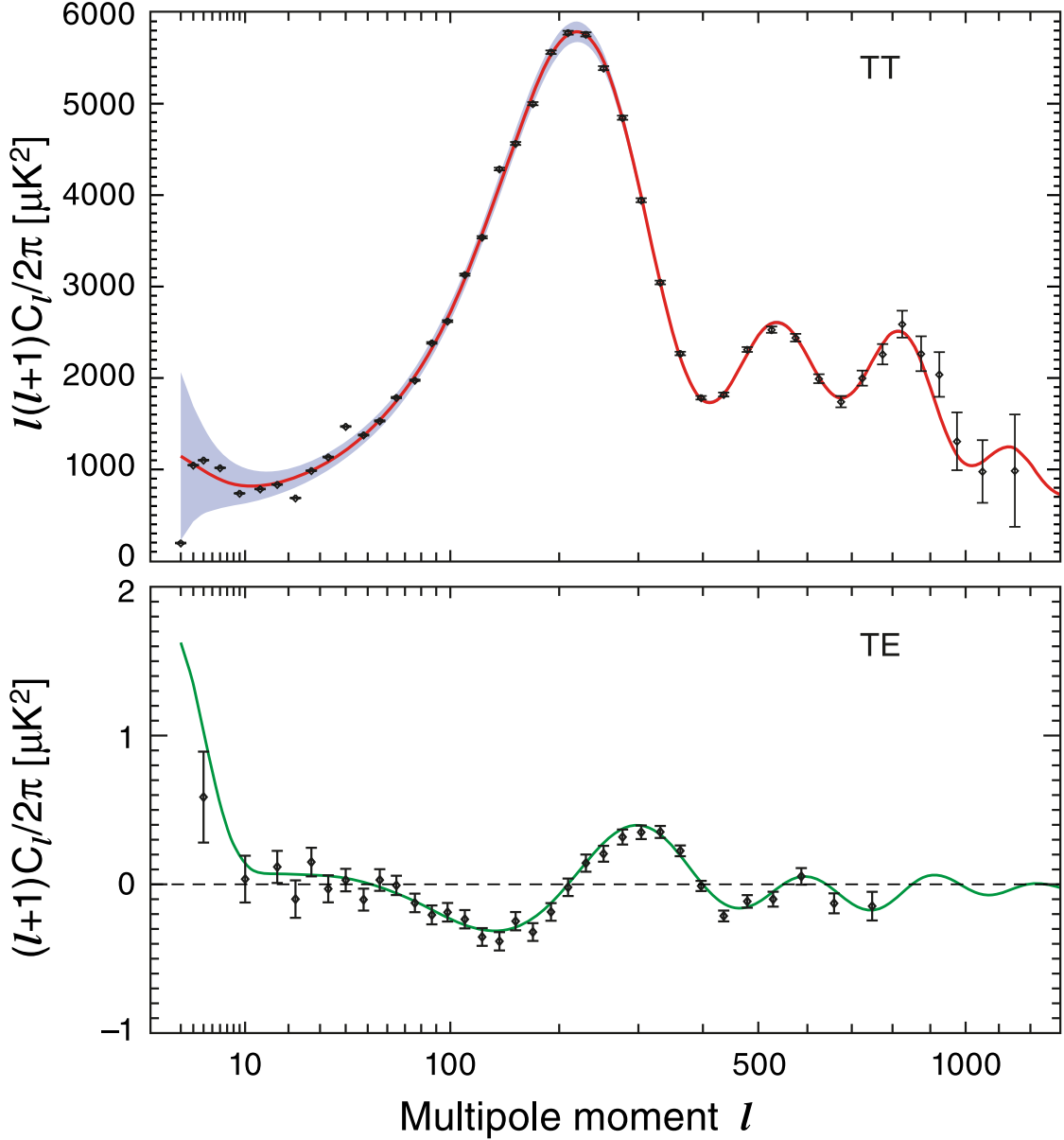}
\caption{ The temperature (TT) and temperature-polarization(TE) power spectra for the seven-year \wmap data set.
The solid lines show the predicted spectrum for the best-fit flat $\Lambda$CDM model. The error bars on the data points
represent measurement errors while the shaded region indicates the uncertainty in the model spectrum arising 
from cosmic variance. \label{fig:final_spectra}The model parameters are: 
\ensuremath{\Omega_bh^2} = \ensuremath{0.02260\pm 0.00053}, 
\ensuremath{\Omega_ch^2} = \ensuremath{0.1123\pm 0.0035},
\ensuremath{\Omega_\Lambda} = \ensuremath{0.728^{+ 0.015}_{- 0.016}},
\ensuremath{n_s} = \ensuremath{0.963\pm 0.012}, 
\ensuremath{\tau} = \ensuremath{0.087\pm 0.014} \ and
\ensuremath{\sigma_8} = \ensuremath{0.809\pm 0.024}. }
\end{figure}

The \wmap data remains one of the cornerstone data sets used for testing the cosmological models and the 
precision measurement of their parameters.
Figure~\ref{fig:final_spectra} displays the binned TT and TE angular power spectra measured from the seven-year 
\wmap data~\citep{larson/etal:prep}, along with the predicted spectrum for the best fit minimal six-parameter flat $\Lambda CDM$ model. 
The overall agreement is excellent, supporting the validity of this model. Table~\ref{tab:best_param} tabulates 
the parameter values for this model using \wmap data alone, and in combination with other data sets. 
Details of the methodology used to determine these values are described in \citet{larson/etal:prep} and
\citet{komatsu/etal:prep}. 

The seven-year \wmap results significantly reduce the uncertainties
for numerous cosmological parameters relative to the five-year results.   The uncertainties in  the  densities 
of baryonic and dark matter are reduced by 10\% and 13\% respectively. When tensor modes are included, 
the upper bound to their amplitude, determined using \wmap data alone, is nearly 20\% lower. By combining \wmap data with 
the latest distance measurements from Baryon Acoustic Oscillations (BAO) in the distribution of galaxies \citep{percival/etal:2009}
and Hubble constant measurements \citep{riess/etal:2009}, the spectral index of the power spectrum of 
primordial curvature perturbations is 
\ensuremath{n_s} = \ensuremath{0.963\pm 0.012}, excluding the 
Harrison-Zel'dovich-Peebles spectrum by more than 3$\sigma$.

The reduced noise obtained by using the seven-year data set yields a better measurement
of the third acoustic peak in the temperature power spectrum. This measurement, when combined
 with external data sets, leads to  better determinations of the total mass of neutrinos, \ensuremath{\sum m_\nu}, 
and the effective number of neutrino species, \ensuremath{N_{\rm eff}}, as presented in Table \ref{tab:best_param}. 
Additionally, when augmented by the small scale CMB anisotropy measurements by
ACBAR \citep{Reichardt/etal:2009} and QUaD \citep{Brown/etal:2009}, this result yields 
a greater that $3\sigma$ detection of the primordial Helium abundance, 
$\ensuremath{Y_{\rm He}} = 0.326 \pm 0.075$, using CMB data alone.
\citet{komatsu/etal:prep} also demonstrate that, with the larger data set, the 
expected radial and azimuthal polarization patterns around hot and cold peaks in the CMB can now be observed directly in pixel-space
by stacking sky map data. In addition, they now detect the Sunyaev-Zel'dovich effect at $\approx 8 \sigma$ at the location of 
known galaxy clusters, as determined by ROSAT.

Finally, \citet{weiland/etal:prep} have  measured the brightness temperature of Jupiter, Saturn, Mars, Uranus and Neptune, 
and five fixed calibrations objects, in all five frequency bands, allowing their use as mm-wave celestial calibration sources traceable to
\wmap's precise calibration.

\section{Summary}
The \wmap observatory has successfully completed seven years of observations with no significant performance
degradation. A full set of sky maps for the seven year data span has been generated and are available for
analyses by the astrophysical community. These maps were generated with an updated masking procedure that
simplifies the map-making procedure and allows creation of a single full-sky noise correlation matrix 
describing the noise correlation over the entire sky for the reduced resolution sky maps. The understanding
of the beam profiles and resulting window functions has been improved with the additional bean profile information
obtained by more observations of Jupiter. The planetary observations, combined with the precise
 absolute calibration of \wmap, were used to measure the brightness temperature of Mars, Jupiter, Saturn, Uranus and Neptune 
for use as calibration sources.

Finally, the  additional data and a better understanding of the instrument's characteristics has 
resulted in tighter constraints on the value of parameters of cosmological models. 
  
\acknowledgments
\wmap is funded by the Science Mission Directorate Office at NASA Headquarters. We acknowledge use of the 
HEALPix~\citep{Gorski/etal:2005} software package. The numerous data products described in this document are available 
from the Legacy Archive for Microwave Background Data Analysis (LAMBDA) \texttt{http://lambda.gsfc.nasa.gov}.

\begin{deluxetable}{lccc}
\tabletypesize{\scriptsize}
\tablecaption{\wmap Seven-year Cosmological Parameter Summary \label{tab:best_param}}
\tablewidth{0pt}
\tablehead{
\colhead{Description} & \colhead{Symbol} & \colhead{\wmap-only} & \colhead{\wmap+BAO+\ensuremath{H_0}}}
\startdata
\multicolumn{4}{c}{Parameters for Standard $\Lambda$CDM Model \tablenotemark{a}} 
    \\[3.0mm]
Age of universe
    &\ensuremath{t_0}
    &\ensuremath{13.75\pm 0.13\ \mbox{Gyr}}
    &\ensuremath{13.75\pm 0.11\ \mbox{Gyr}}
    \\[1.5mm]
Hubble constant
    &\ensuremath{H_0}
    &\ensuremath{71.0\pm 2.5\ \mbox{km/s/Mpc}}
    &\ensuremath{70.4^{+ 1.3}_{- 1.4}\ \mbox{km/s/Mpc}}
    \\[1.5mm]
Baryon density
    &\ensuremath{\Omega_b}
    &\ensuremath{0.0449\pm 0.0028}
    &\ensuremath{0.0456\pm 0.0016}   
    \\[1.5mm]
Physical baryon density
    &\ensuremath{\Omega_bh^2}
    &\ensuremath{0.02258^{+ 0.00057}_{- 0.00056}}
    &\ensuremath{0.02260\pm 0.00053}
    \\[1.5mm]
Dark matter density
    &\ensuremath{\Omega_c}
    &\ensuremath{0.222\pm 0.026}
    &\ensuremath{0.227\pm 0.014}
    \\[1.5mm]
Physical dark matter density
    &\ensuremath{\Omega_ch^2}
    &\ensuremath{0.1109\pm 0.0056}
    &\ensuremath{0.1123\pm 0.0035}
    \\[1.5mm]
Dark energy density
    &\ensuremath{\Omega_\Lambda}
    &\ensuremath{0.734\pm 0.029}
    &\ensuremath{0.728^{+ 0.015}_{- 0.016}}
    \\[1.5mm]
Curvature fluctuation amplitude, $k_0=0.002$ Mpc$^{-1}$ \tablenotemark{b}
    &\ensuremath{\Delta_{\cal R}^2}
    &\ensuremath{(2.43\pm 0.11)\times 10^{-9}}
    &\ensuremath{(2.441^{+ 0.088}_{- 0.092})\times 10^{-9}}
    \\[1.5mm]
Fluctuation amplitude at $8h^{-1}$ Mpc
    &\ensuremath{\sigma_8}
    &\ensuremath{0.801\pm 0.030}
    &\ensuremath{0.809\pm 0.024}
    \\[1.5mm]
Scalar spectral index
    &\ensuremath{n_s}
    &\ensuremath{0.963\pm 0.014}
    &\ensuremath{0.963\pm 0.012}
    \\[1.5mm]
Redshift of matter-radiation equality
    &\ensuremath{z_{\rm eq}}
    &\ensuremath{3196^{+ 134}_{- 133}}  
    &\ensuremath{3232\pm 87}    
    \\[1.5mm]
Angular diameter distance to matter-radiation eq.\tablenotemark{c}
    &\ensuremath{d_A(z_{\rm eq})}
    &\ensuremath{14281^{+ 158}_{- 161}\ \mbox{Mpc}}
    &\ensuremath{14238^{+ 128}_{- 129}\ \mbox{Mpc}}
    \\[1.5mm]
Redshift of decoupling
    &\ensuremath{z_{*}}
    &\ensuremath{1090.79^{+ 0.94}_{- 0.92}}
    &\ensuremath{1090.89^{+ 0.68}_{- 0.69}}
    \\[1.5mm]
Age at decoupling
    &\ensuremath{t_{*}}
    &\ensuremath{379164^{+ 5187}_{- 5243}\ \mbox{yr}}
    &\ensuremath{377730^{+ 3205}_{- 3200}\ \mbox{yr}}
    \\[1.5mm]
Angular diameter distance to decoupling \tablenotemark{c,d}
    &\ensuremath{d_A(z_{*})}
    &\ensuremath{14116^{+ 160}_{- 163}\ \mbox{Mpc}}
    &\ensuremath{14073^{+ 129}_{- 130}\ \mbox{Mpc}}
    \\[1.5mm]
Sound horizon at decoupling \tablenotemark{d}
    &\ensuremath{r_s(z_*)}
    &\ensuremath{146.6^{+ 1.5}_{- 1.6}\ \mbox{Mpc}}
    &\ensuremath{146.2\pm 1.1\ \mbox{Mpc}}
    \\[1.5mm]
Acoustic scale at decoupling \tablenotemark{d}
    &$l_A(z_*)$
    &\ensuremath{302.44\pm 0.80}
    &\ensuremath{302.40\pm 0.73}
    \\[1.5mm]
Reionization optical depth
    &\ensuremath{\tau}
    &\ensuremath{0.088\pm 0.015}  
    &\ensuremath{0.087\pm 0.014}    
    \\[1.5mm]
Redshift of reionization
    &\ensuremath{z_{\rm reion}}
    &\ensuremath{10.5\pm 1.2}  
    &\ensuremath{10.4\pm 1.2}    
    \\[1.5mm]
\multicolumn{4}{c}{Parameters for Extended Models \tablenotemark{e}} 
    \\[3.0mm]
Total density \tablenotemark{f}
    &\ensuremath{\Omega_{\rm tot}}
    &\ensuremath{1.080^{+ 0.093}_{- 0.071}}
    &\ensuremath{1.0023^{+ 0.0056}_{- 0.0054}}
    \\[1.5mm]
Equation of state \tablenotemark{g}
    &\ensuremath{w}
    &\ensuremath{-1.12^{+ 0.42}_{- 0.43}}
    & -0.980 $\pm$ 0.053
    \\[1.5mm]
Tensor to scalar ratio, $k_0=0.002$ Mpc$^{-1}$ \tablenotemark{b,h}
    &\ensuremath{r}
    &\ensuremath{< 0.36\ \mbox{(95\% CL)}}
    &\ensuremath{< 0.24\ \mbox{(95\% CL)}}
    \\[1.5mm]
Running of spectral index, $k_0=0.002$ Mpc$^{-1}$ \tablenotemark{b,i}
    &\ensuremath{dn_s/d\ln{k}}
    &\ensuremath{-0.034\pm 0.026}
    &\ensuremath{-0.022\pm 0.020}
    \\[1.5mm]
Neutrino density \tablenotemark{j}
    &\ensuremath{\Omega_\nu h^2}
    &\ensuremath{< 0.014\ \mbox{(95\% CL)}}
    &\ensuremath{< 0.0062\ \mbox{(95\% CL)}}
    \\[1.5mm]
Neutrino mass \tablenotemark{j}
    &\ensuremath{\sum m_\nu}
    &\ensuremath{< 1.3\ \mbox{eV}\ \mbox{(95\% CL)}}
    &\ensuremath{< 0.58\ \mbox{eV}\ \mbox{(95\% CL)}}
    \\[1.5mm]
Number of light neutrino families \tablenotemark{k}
    &\ensuremath{N_{\rm eff}}
    &\ensuremath{> 2.7\ \mbox{(95\% CL)}}
   &\ensuremath{4.34^{+ 0.86}_{- 0.88}}
    \\[1.5mm]
\enddata
\tablenotetext{a}{The parameters reported in the first section assume the 6
parameter flat $\Lambda$CDM model, first using \wmap\ data only
\citep{larson/etal:prep}, then using \wmap+BAO+\ensuremath{H_0} data
\citep{komatsu/etal:prep}. The \ensuremath{H_0} data consists of a Gaussian prior on the present-day value of the Hubble
constant, \ensuremath{H_0} = 74.2 $\pm$ 3.6 km $\rm s^{-1}$ $\rm Mpc^{-1}$\citep{riess/etal:2009}, while the BAO priors
on the distance ratio $r_s(z_d)/D_{\rm V}(z)$ at $ z = 0.2,~0.3$ are obtained from the Sloan Digital Sky Survey Data 
Release 7~\citep{percival/etal:2009}. Uncertainties are 68\% CL unless otherwise noted.}
\tablenotetext{b}{$k=0.002$ Mpc$^{-1}$  $\longleftrightarrow$  $l_{\rm eff} \approx 30$.}
\tablenotetext{c}{Comoving angular diameter distance.}
\tablenotetext{d}{$l_A(z_*) \equiv \pi \, \ensuremath{d_A(z_{*})} \, \ensuremath{r_s(z_*)}^{-1}$.}
\tablenotetext{e}{The parameters reported in the second section place limits on
deviations from the $\Lambda$CDM model, first using \wmap\ data only
\citep{larson/etal:prep}, then using \wmap+BAO+\ensuremath{H_0} data
\citep{komatsu/etal:prep}, except as noted otherwise.  A complete listing of all parameter values and
uncertainties for each of the extended models studied is available on LAMBDA.}
\tablenotetext{f}{Allows non-zero curvature, $\Omega_k \ne 0$.}
\tablenotetext{g}{Allows $w \ne -1$, but assumes $w$ is constant and $\Omega_k = 0$. The value in the last column is obtained from a 
combination of \wmap +BAO data and luminosity distance information obtained from the ``constitution'' SNe data set~\citep{hicken/etal:2009}
 using the methodology described in~\citet{komatsu/etal:prep}} 
\tablenotetext{h}{Allows tensors modes but no running in scalar spectral index.}
\tablenotetext{i}{Allows running in scalar spectral index but no tensor modes.}
\tablenotetext{j}{Allows a massive neutrino component, $\Omega_{\nu} \ne 0$.}
\tablenotetext{k}{Allows $N_{\rm eff}$ number of relativistic species, with the prior $0< \ensuremath{N_{\rm eff}} <10$.}
\end{deluxetable}


\clearpage

\begin{thebibliography}{25}
\expandafter\ifx\csname natexlab\endcsname\relax\def\natexlab#1{#1}\fi

\bibitem[{Barrett et~al.(1994)}]{templates}
Barrett, R., et~al. 1994, Templates for the Solution of Linear Systems:
  Building Blocks for Iterative Methods, 2nd Edition (Philadelphia, PA: SIAM)

\bibitem[{{Bennett} et~al.(2003{\natexlab{a}})}]{bennett/etal:2003b}
{Bennett}, C.~L., et~al. 2003{\natexlab{a}}, \apjs, 148, 1

\bibitem[{{Bennett} et~al.(2003{\natexlab{b}})}]{bennett/etal:2003}
---. 2003{\natexlab{b}}, \apj, 583, 1

\bibitem[{{Bennett} et~al.(2010)}]{bennett/etal:prep}
{Bennett}, C.~L. et~al. 2010, in preparation

\bibitem[{{Brown} et~al.(2009)}]{Brown/etal:2009}
{Brown}, M.~L., et~al. 2009, \apj, 705, 978

\bibitem[{{Gold} et~al.(2010)}]{gold/etal:prep}
{Gold}, B. et~al. 2010, in preparation

\bibitem[{Gorski et~al.(2005)Gorski, Hivon, Banday, Wandelt, Hansen, Reinecke,
  \& Bartlemann}]{Gorski/etal:2005}
Gorski, K.~M., Hivon, E., Banday, A.~J., Wandelt, B.~D., Hansen, F.~K.,
  Reinecke, M., \& Bartlemann, M. 2005, \apj, 622, 759

\bibitem[{{Hicken} et~al.(2009){Hicken}, {Wood-Vasey}, {Blondin}, {Challis},
  {Jha}, {Kelly}, {Rest}, \& {Kirshner}}]{hicken/etal:2009}
{Hicken}, M., {Wood-Vasey}, W.~M., {Blondin}, S., {Challis}, P., {Jha}, S.,
  {Kelly}, P.~L., {Rest}, A., \& {Kirshner}, R.~P. 2009, \apj, 700, 1097

\bibitem[{{Hill} et~al.(2009)}]{hill/etal:2009}
{Hill}, R.~S., et~al. 2009, \apjs, 180, 246

\bibitem[{Hinshaw et~al.(2003)}]{hinshaw/etal:2003d}
Hinshaw, G. et~al. 2003, Astrophys. J. Suppl., 148, 63

\bibitem[{{Hinshaw} et~al.(2009)}]{hinshaw/etal:2009}
{Hinshaw}, G., et~al. 2009, \apjs, 180, 225

\bibitem[{{Hivon} et~al.(2002){Hivon}, {G{\' o}rski}, {Netterfield}, {Crill},
  {Prunet}, \& {Hansen}}]{hivon/etal:2002}
{Hivon}, E., {G{\' o}rski}, K.~M., {Netterfield}, C.~B., {Crill}, B.~P.,
  {Prunet}, S., \& {Hansen}, F. 2002, \apj, 567, 2

\bibitem[{{Jarosik} et~al.(2003)}]{jarosik/etal:2003}
{Jarosik}, N., et~al. 2003, \apjs, 145, 413

\bibitem[{{Jarosik} et~al.(2007)}]{jarosik/etal:2007}
---. 2007, \apjs, 170, 263

\bibitem[{{Komatsu} et~al.(2010)}]{komatsu/etal:prep}
{Komatsu}, E. et~al. 2010, in preparation

\bibitem[{{Larson} et~al.(2010)}]{larson/etal:prep}
{Larson}, D. et~al. 2010, in preparation

\bibitem[{{Limon} et~al.(2010)}]{limon/etal:prep}
{Limon}, M., et~al. 2010, Wilkinson Microwave Anisotropy Probe ({\sl WMAP}):
  Seven Year Explanatory Supplement,
  \texttt{http://lambda.gsfc.nasa.gov/data/map/doc/MAP\_supplement.pdf}

\bibitem[{{Mather} et~al.(1999){Mather}, {Fixsen}, {Shafer}, {Mosier}, \&
  {Wilkinson}}]{mather/etal:1999}
{Mather}, J.~C., {Fixsen}, D.~J., {Shafer}, R.~A., {Mosier}, C., \&
  {Wilkinson}, D.~T. 1999, \apj, 512, 511

\bibitem[{{Page} et~al.(2003{\natexlab{a}})}]{page/etal:2003b}
{Page}, L., et~al. 2003{\natexlab{a}}, \apjs, 148, 39

\bibitem[{{Page} et~al.(2003{\natexlab{b}})}]{page/etal:2003}
---. 2003{\natexlab{b}}, \apj, 585, 566

\bibitem[{{Page} et~al.(2007)}]{Page/etal:2007}
---. 2007, \apjs, 170, 335

\bibitem[{{Percival} et~al.(2009)}]{percival/etal:2009}
{Percival}, W.~J., et~al. 2009, \mnras, 1741

\bibitem[{{Reichardt} et~al.(2009)}]{Reichardt/etal:2009}
{Reichardt}, C.~L., et~al. 2009, \apj, 694, 1200

\bibitem[{{Riess} et~al.(2009)}]{riess/etal:2009}
{Riess}, A.~G., et~al. 2009, \apj, 699, 539

\bibitem[{{Weiland} et~al.(2010)}]{weiland/etal:prep}
{Weiland}, J.~L. et~al. 2010, in preparation

\end{thebibliography}

\end{document}